\documentclass[11pt, a4paper]{article}
\usepackage[tableposition=top]{caption}
\usepackage[onehalfspacing]{setspace}
\usepackage[top=70pt,bottom=50pt,left=80pt,right=70pt]{geometry}

\usepackage{fancyhdr}
\usepackage{color}
\usepackage{amssymb,amsmath,amsthm,bbm,mathrsfs,ulem}
\usepackage{latexsym} 
\usepackage{graphicx} 
\usepackage{verbatim}
\usepackage{nicefrac} 
\usepackage{hyperref}
\usepackage[latin1]{inputenc} 
\usepackage{doc,natbib} 
\usepackage[dvipsnames, table]{xcolor}

\graphicspath{{./}}
\bibpunct{(}{)}{;}{a}{,}{,} 

\newtheorem{theo}{Theorem}[section]
\newtheorem{coro}[theo]{Corollary}
\newtheorem{prop}[theo]{Proposition}

\newtheorem{lemm}[theo]{Lemma}

\newtheorem{algo}[theo]{Algorithm}

\newcommand{\ew}{\mathbb{E}}     
\newcommand{\va}{\mathbb{V}\!ar} 
\newcommand{\pr}{\mathbb{P}}     
\newcommand{\rn}{\mathbb{R}}     
 
\newcommand{\nn}{\mathbb{N}}		

\DeclareMathOperator*{\argmax}{argmax}

\newcommand\blfootnote[1]{%
	\begingroup
	\renewcommand\thefootnote{}\footnote{#1}%
	\addtocounter{footnote}{-1}%
	\endgroup
}

\title{Multiscale change point detection\\ via gradual bandwidth adjustment\\ in moving sum processes}
\author{
	{Tijana Levajkovi\'c and Michael Messer\footnote{corresponding author}}\\[1ex]
	{Vienna University of Technology}\\
    }
\date{}

\begin{document}
	\maketitle
	
	\begin{abstract}
		A method for the detection of changes in the expectation in univariate sequences is provided. Moving sum processes (MOSUM) are studied. These rely on the selection of a tuning bandwidth. Here, a framework to overcome bandwidth selection is presented \--- the bandwidth adjusts gradually. For that, MOSUM are made dependent on both time and the bandwidth: the domain becomes a triangle. On the triangle, paths are constructed which systematically lead to change points. An algorithm is provided that estimates change points by subsequent consideration of paths. Strong consistency for the number and location of change points is shown. Simulation studies corroborate estimation precision and reveal competitiveness with state of the art change point detection methods. A companion $\texttt{R}$-package \texttt{mscp} is made available on CRAN.
		\vspace{0.5em}
		
	\end{abstract}	
	

\section{Introduction}\label{sect:introduction}

We contribute to the field of change point detection in stochastic sequences.
Change point detection applies in various research areas, e.g., climatology \citep{Reeves2007}, 
speech recognition \citep{Rybach2009}, oceanography \citep{Killick2010}, neuroimaging \citep{Aston2012}, virology \citep{Kass2012} etc. 

\begin{figure}[h!]
	\centering  
	\includegraphics[width=0.8\textwidth,angle=0]{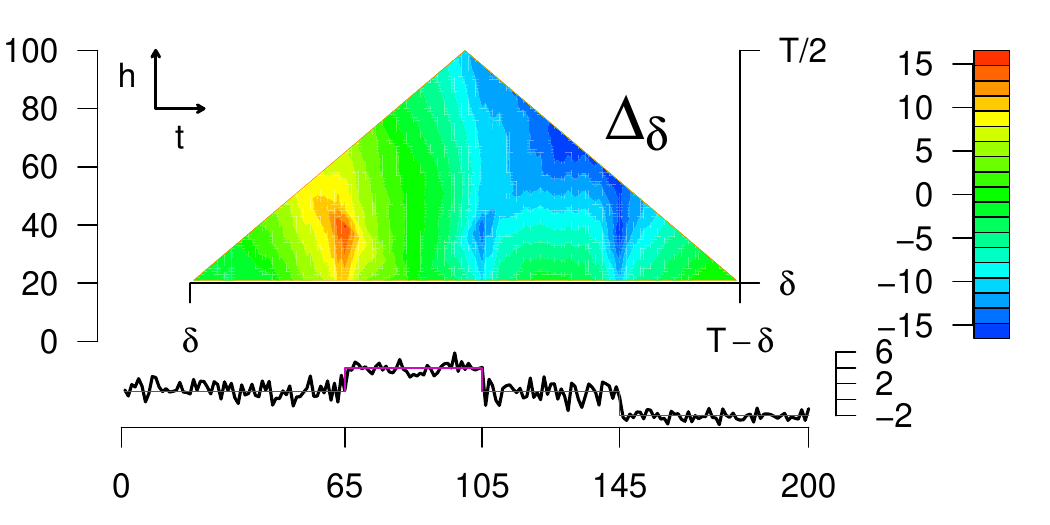}
	\caption{Bottom: Process via $N(\mu,\sigma^2)$ RVs, with $T= 200$, $C=\{65,105,145\}$, $\mu=1,4,1,-2$ (pink), $\sigma=1,0.8,1,0.5$. Top: $D_{t,h}$ with $(t,h)\in \Delta_\delta$ for $\delta=20$.}
	\label{fig:expl1a}
\end{figure}

We consider $T$ univariate and independent random variables (RVs)  $X_1,\ldots, X_T$, that are piecewise identically distributed, with existing $(2+p)$-th moments ($p>0$), without parametric assumptions. Multiple change points in expectation form a set $C$. See Figure \ref{fig:expl1a} (bottom) for an example with $T=200$, $X_i\sim N(\mu,\sigma^2)$, three change points $C=\{65,105,145\}$, and thus four sections with parameters $\mu=1,4,1,-2$ and $\sigma=1,0.8,1,0.5$, i.e., changes in $\sigma$ may additionally occur when $\mu$ changes. 

There is extensive literature that covers changes in expectation, e.g., methods based on likelihood ratios \citep{Fang2020,Gombay1994}, empirical processes \citep{Holmes2013,Horvath2007}, $U$-statistics \citep{Doering2010,Gombay2002,Horvath2005}, least-squares \citep{Harchaoui2010,Lavielle2000} and many more. We mention methodology based on CUSUM-statistics, e.g., by \citep{Berkes2006,Dehling2017,Hinkley1971,Page1954}. For a general overview of change point methods see the textbooks of \citet{Chen2000,Brodsky2017,Csorgo1997}. In this paper, we aim to tackle multiple change points that may occur on different time scales, as considered e.g., in \citet{fryz2014,matteson2014,Pein2016,Spokoiny2009}. We study MOSUM, see e.g., \citet{Antoch1999,Chu1995,Huskova2001,Steinebach1995}. For that we select a window size (bandwidth) $h\in \{1,\ldots,\lfloor T/2\rfloor\}$ and define MOSUM $(D_{t,h})_{t}$ for index $t = h,\ldots,T-h$: for every time $t$ consider two adjacent windows of size $h$, left $\{t-h+1,\ldots,t\}$ (index $\ell$) and right $\{t+1,\ldots,t+h\}$ (index $r$), and set
\begin{align}\label{Dintro}
D_{t,h} := \sqrt{h}\cdot\frac{\hat \mu_r - \hat\mu_\ell}{(\hat\sigma_r^2+\hat\sigma_\ell^2)^{1/2}}
\end{align}
where $\hat \mu_j$ and $\hat \sigma_j^2$ denote the mean and empirical variance of the RVs whose indices lie in the windows, $j\in\{\ell,r\}$. 
$D_{t,h}$ is Welch's $t$-statistic for two samples of size $h$. We typically find $|D_{t,h}|\approx 0$ if no change is involved, but $|D_{t,h}|>0$ if there is a change nearby, $t\approx c\in C$. Thus, change point estimates may be obtained by argmax-estimation, see e.g., \citet{Eichinger2018}. 

A major challenge lies in the choice of the window size $h$. An $h$ small enough is sensitive to rapid changes when it does not overlap subsequent change points,  while a larger $h$ improves detection power of small effects as more RVs are evaluated. But note that $h$ too large may result in overlap of subsequent changes and thus in an estimation bias, or even a failure of detection at all. In order to account for change points that occur on multiple time scales, including rapid changes as well as small effects, methods that combine multiple windows were proposed, see e.g., \citet{Cho2020,Messer2021}. They work in two steps: first change point candidates are generated for every single $h$, and afterwards all sets are merged giving final estimates. Despite improvements, the methods demand the selection of a window set that best accounts for the location of unknown change points. 

The aim of this paper is to provide a MOSUM framework that overcomes window selection, but nevertheless exploits multiple windows to address change point occurrences on multiple time scales, denoted \textit{multi-scale change point detection} algorithm (MSCP),  see Algorithm \ref{algo:estimation}.  For that we extend the MOSUM perspective: instead of considering $(D_{t,h})_t$ as a process of time $t$ only, we let it depend on both $t$ and $h$, i.e., $(D_{t,h})_{(t,h)}$, while the indices $(t,h)$ lie in a  triangle $\Delta_\delta\subset\rn^2$, see Figure \ref{fig:expl1a} (top). We define $\Delta_\delta:= \{ (t,h) \;|\; t\in T_h, h\in H_\delta\}$, for which we consider first $\delta\in \{1,2,\ldots, \lfloor T/2 \rfloor\}$ a fixed minimal window, while $\lfloor \cdot \rfloor$ denotes the floor function, second $H_\delta:=[\delta,T/2]$ a window interval, and third $T_h:=[h,T-h]$ a time interval. Note, 
$\Delta_\delta$ is a right-angled and isosceles triangle, and the hypotenuse is oriented as the lower edge and refers to the smallest window $h=\delta$. A higher horizontal slice refers to a larger $h$, and the upper vertex describes the largest $h=\lfloor T/2\rfloor$.

The triangular structure follows from shrinkage of possible $t$-indices $h,\ldots,T-h$ when $h$ increases. In Figure \ref{fig:expl1a} $D_{t,h}$ is color-coded with $D_{t,h}\approx 0$ green,  $>0$ red, and $<0$ blue. At $c_1=65$ there is an increase in $\mu$ and thus $D_{t,h}>0$, while at $c_2=105$ and $c_3=145$ a decrease in $\mu$ yields $D_{t,h}<0$, at least when $h$ is not too large ($h<40$) such that only a single $c_u$ is overlapped. The upper part of $\Delta_\delta$ refers to larger $h$ that result in an overlap of multiple change points. The area between $c_1$ and $c_2$ is green also for $h$ large, at $(t,h)\approx(85,70)$, because the parameters in the first and third section coincide an thus the effects cancel out. In contrast, the area between $c_ 2$ and $c_3$ is dark blue for $h$ large, at $(t,h)\approx(120,70)$, as both changes at $c_2$ and $c_3$ are negative, which amplifies the effect and as a consequence simple argmax estimation would be flawed. Importantly, note that for smaller $h$ large values of $|D_{t,h}|$ concentrate around $c_u\in C$.

\begin{figure}[h!]
	\centering  
	\includegraphics[width=0.7\textwidth,angle=0]{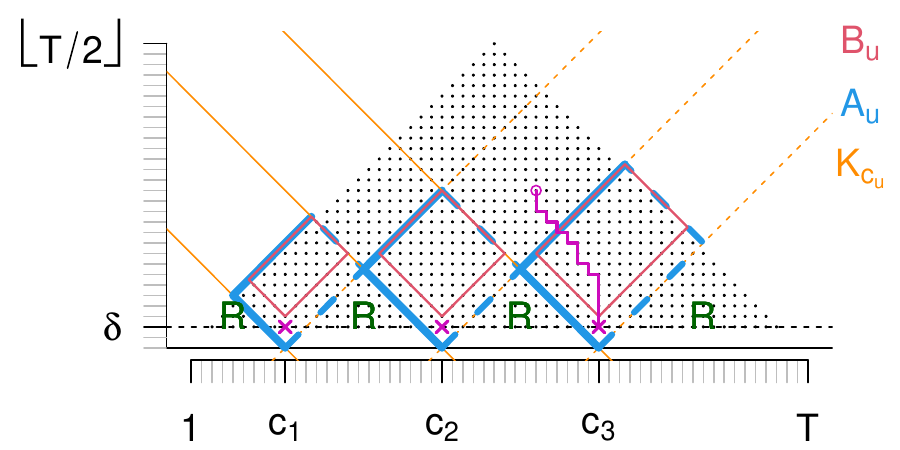}
	\caption{Representation of areas of attraction $A_u$ (blue), inner sets $B_u$ (red), cones $K_{c_u}$ (orange) for three change points $c_u$, the remainder $R$ (green), and a zigzag-path (magenta).}
	\label{fig:attraction}
\end{figure}

MSCP subsequently acts on subsets of $\Delta_\delta$ by locally exploiting $(D_{t,h})_{(t,h)}$. The key ingredient is the construction of a \textit{zigzag-path}, see Figure \ref{fig:attraction} (magenta): given a starting value $(t_s,h_s)$ (pink circle), the path leads towards the lower edge of $\Delta_\delta$ to some point $(\hat c,\delta)$, and $\hat c$ functions as a change point estimate. The path evolves according to stepwise local argmax-estimation: in each instance the path moves one step downwards, i.e., $h$ switches to $h-1$ non-randomly. Then the path moves either one step left or right, or it stays, i.e., $t$ switches to some value in $\{t-1,t,t+1\}$, while the choice falls on the $t$-maximizer of $|D_{t,h-1}|$. 

\begin{figure}[h!]
	\centering  
	\includegraphics[width=0.7\textwidth,angle=0]{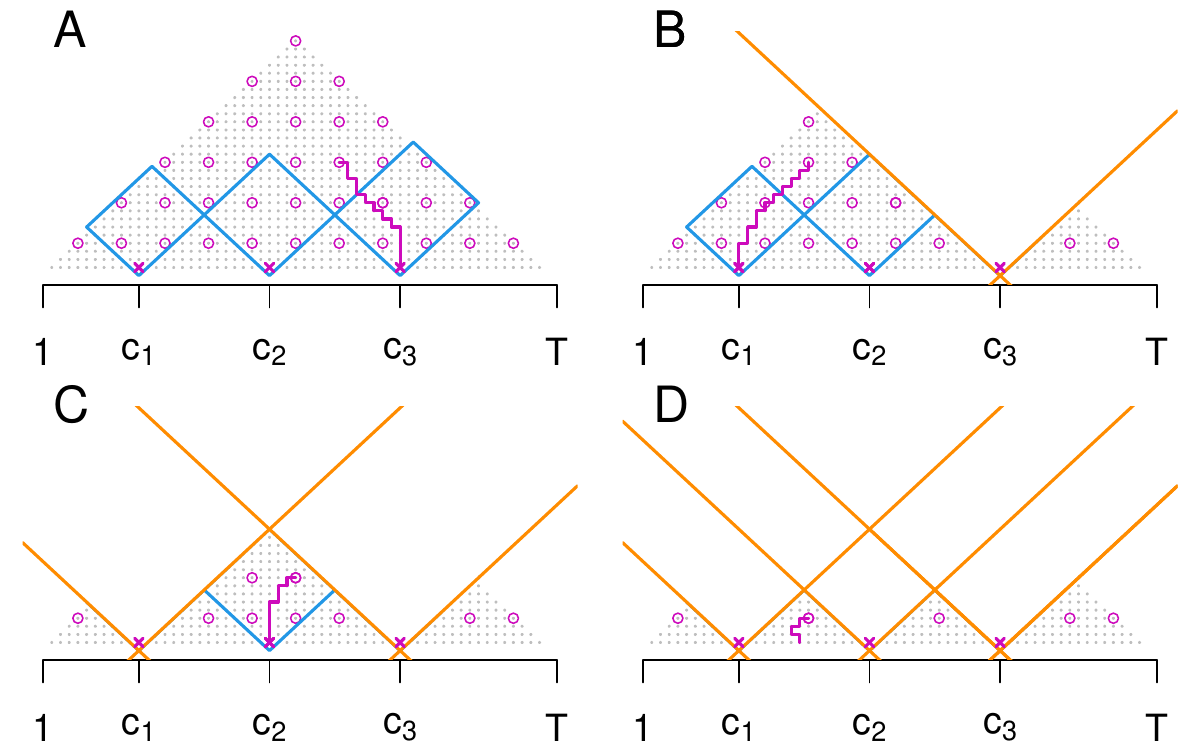}
	\caption{Schematic representation of MSCP. It is $C=\{c_1,c_2,c_3\}$. Subsequently $c_3$, $c_1$ and $c_2$ are estimated, and then MSCP breaks.}
	\label{fig:cutout}
\end{figure}

The idea of MSCP is the following, see also Figure \ref{fig:cutout}: given a set $S\subset\Delta_\delta$ of possible starting values (pink circles), the one maximizing $h^{-1/2}\cdot |D_{t,h}|$ (i.e., the strongest signal to noise ratio) is the first starting point considered. Then its path delivers the first change point estimate $\hat c$. In order to avoid false positives, a breaking criterion is evaluated, which is computed from $D_{t,h}$ on the path. If breaking is not demanded, then $\hat c$ is accepted. In order to avoid multiple detections of the same change point, all elements of $S$, whose paths could lead to $\hat c$, are deleted from $S$ ('cut out cone'). Then estimation restarts, and iteratively change points are detected until the breaking criterion applies. In Figure \ref{fig:cutout} the algorithm estimates $c_3$, $c_1$ and $c_2$, and then the breaking criterion applies for the next  candidate.

The main result states strong consistency of MSCP for both the number and location of estimated change points. The crucial technical foundation 
relies on the path behavior: if the starting point $(t_s,h_s)$ lies high enough on $\Delta_\delta$, i.e., $h_s$ large, then at some point the path enters a subset of $\Delta_\delta$ that we call the \textit{area of attraction} $A_u$ of a change point $c_u\in C$, see Figure \ref{fig:attraction} (blue), and from this on it is systematically pushed towards the $c_u$.  

\begin{figure}[h!]
	\centering  
	\includegraphics[width=0.7\textwidth,angle=0]{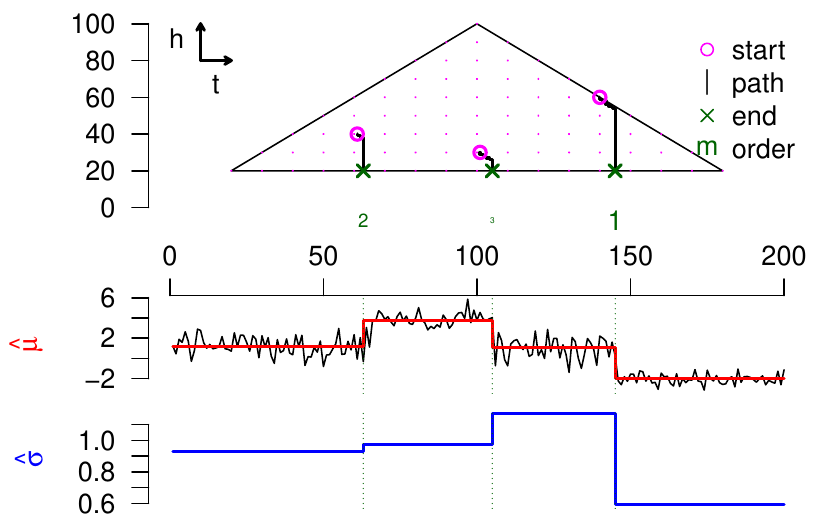}
	\caption{Plotting routine of the $\texttt{R}$-package \texttt{mscp}. It is $\hat C =\{63, 105, 145 \}$.}
	\label{fig:expl1b}
\end{figure}

For the process	in Figure \ref{fig:expl1a} MSCP yields estimates $145,63$ and $105$ depicted in Figure \ref{fig:expl1b}, i.e., the number is correct, the smallest estimate at $63$ is close to $c_1=65$ and the other two estimates even hit the true $c_2=105$ and $c_3=145$. MSCP is made available in the $\texttt{R}$-package $\texttt{mscp}$ on CRAN \citet{CRANMSCP},  which includes a summary and plotting routine while the latter created Figure \ref{fig:expl1b}: the top panel shows $\Delta_\delta$ including $S$, and the paths constructed with detection orders indicated by integers (green). The middle panel gives the process and the means of the RVs in the detected sections (red), and the bottom shows the empirical variances therein (blue).

We strengthen three important upsides of MSCP: first, the method is non-parametric with weak distributional assumptions and thus allows to analyze a high variety of data. Robustness against additional changes in variance could be helpful in practice, as e.g., an increase in the mean might be accompanied with an increase in volatility. Second, the problem of window selection is overcome. Third, the gradual window adjustment improves the simultaneous detection of change points on multiple time scales and of different signal to noise ratios.

Estimation precision is corroborated by simulation studies investigating  different change point scenarios and effect sizes. Various distributions are considered, including normal, gamma, Poisson and binomial, and their combinations. A comparison with state of the art methods reveals strong competitiveness of MSCP.  

We mention that MSCP could be generalized generically, i.e., extended for the detection of other quantities than $\mu$, e.g., higher moments, changes in slope, etc. For that, on the one hand both MOSUM $D_{t,h}$ in (\ref{Dintro}) and also model formulation needed adjustment, e.g., by considering a difference of empirical higher moments or estimated slopes. On the other hand, the intrinsic structure of MSCP, i.e., the successive construction of paths on $\Delta_\delta$ would be maintained. 

The paper is organized as follows: In Section \ref{sect:model_and_stats} we specify the model, $\Delta_\delta$ and $D_{t,h}$. In Section \ref{sect:D_asymptotics} we study properties of $(D_{t,h})_{(t,h)\in\Delta_\delta}$ which are used for change point detection. In Section \ref{sect:cpd} we construct the zigzag-path and show that it yields proper estimates. We introduce the set of starting points $S$, define MSCP, and state consistency. In Section \ref{sect:practical} we discuss the tuning parameters ($\delta$, $S$ and the breaking criterion), present the simulations, and  discuss a data example where MSCP segments a nucleotide sequence of a human genome. Proofs are given in the 
Appendix \ref{sect:appendix}.

\section{The change point model and MOSUM on the triangle}\label{sect:model_and_stats}
\paragraph{The model $\mathcal M$}
We fix $T\in\nn\backslash\{0,1\}$ and call $\{1,\ldots,T\}$ the observation regime. We consider a subset $C \subset \{1,\ldots,T-1\}$ of cardinality $|C|$ with ordered elements $c_1 < c_2 \ldots < c_{|C|}$. We call $c_u$ the $u$-th change point and $C$ the set of change points. $C$ is treated fixed but unknown. Set $c_0:=0$ and $c_{|C|+1}:=T$, and the minimal distance of adjacent change points $\delta_{C}:=\min_{u=1,\ldots,|C|+1} (c_{u}-c_{u-1})$.

Let $(\Omega,\mathcal A,\pr)$ be a probability space and $p>0$. We consider a triangular scheme: let  $(Z_{u,i})_{u,i}$ for $u\in\{1,2,\ldots,|C|+1\}$ and $i=1,2,\ldots$ be independent RVs in $\mathscr L^{2+p}(\Omega,\mathcal A,\pr)$ with $\ew[Z_{u,i}]=0$ and $\va(Z_{u,i})=1$, and let  $(Z_{u,i})_{i=1,2\ldots}$ be an i.i.d.~sequence for each $u$.
Further let $\mu_1,\mu_2,\ldots,\mu_{|C|+1}\in \rn$ with $\mu_u\not=\mu_{u+1}$, and  $\sigma_1^2,\sigma_2^2,\ldots,\sigma_{|C|+1}^2$ positive. For $n\in\{1,2,\ldots\}$ set 
\begin{align}\label{model} 
X_i := \sum_{u=1}^{|C|+1}
(\mu_u + \sigma_u Z_{u,i}) \cdot
\mathbbm{1}_{\{nc_{u-1}+1,\ldots,nc_u\}}(i), \quad i\in \{1, 2, \dots, nT\}.
\end{align}
Dependence of $X_i$ on $n$ is suppressed for simplicity. The process $\mathbf X:=(X_i)_{i=1,2,\ldots,nT}$ has $|C|$ change points. In the section from $nc_{u-1}+1$ to $nc_{u}$ it has expectation $\mu_{u}$ and variance $\sigma_u^2$, which are then changing to $\mu_{u+1}$ and $\sigma_{u+1}^2$. 
The case $n=1$ is considered the real-time scenario. Throughout, asymptotics are studied letting $n\to\infty$, i.e., the observation regime and all change points increase linearly. Consequently, the relative change point location $nc_u/nT$ is constant over $n$. The formulation of $n$ in (\ref{model}) facilitates to track scaling $n^{\vartheta}$ for different $\vartheta$ in the course of the article.
Given $T$, the set of processes $\mathbf X$ constitutes the model $\mathcal M$.

For all $u$, we set the first two moments $\mu_u^{\langle 1 \rangle}:=\mu_u$ and $\mu_u^{\langle 2 \rangle}:= \sigma_u^2 + \mu_u^2$ and the centered moments $\mu_u^{\{1\}}=0$ and $\mu_u^{\{2\}}:=\sigma_u^2$. If $C=\emptyset$, we set
$\mu:=\mu^{\langle 1 \rangle}:=\mu_1^{\langle 1 \rangle}$, $\mu^{\langle 2 \rangle} := \mu_1^{\langle 2 \rangle}$ and $\sigma^2:=\sigma_1^2:=\mu_1^{\{2\}}$. 

\begin{figure}[h!]
	\centering  
	\includegraphics[width=0.2\textwidth,angle=0]{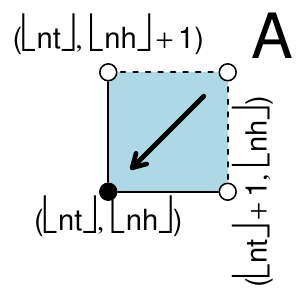}
	\includegraphics[width=0.4\textwidth,angle=0]{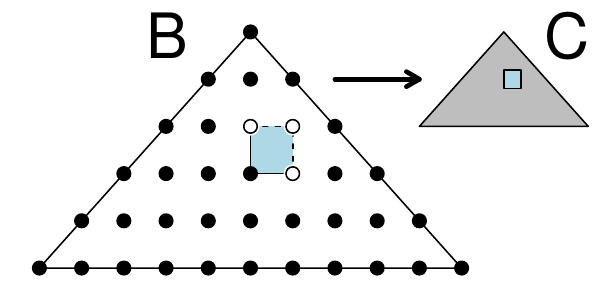}
	\caption{Process construction. The floor-functions (A) in $I_j^{(n)}$, $j\in \{l,r\}$ for both $\lfloor nt \rfloor$ and $\lfloor nh \rfloor$ appear within a factor-$n$-enlarged triangle (B), which is rescaled to $\Delta_\delta$ (C).}	
	\label{fig:cadlag}
\end{figure}

\paragraph{Local estimators and MOSUM} For $(t,h)\in \Delta_\delta$, see Section \ref{sect:introduction}, we consider indices interpreted a left and right window $I_\ell^{(n)}:=\{\lfloor nt\rfloor - \lfloor nh\rfloor+1,\ldots, \lfloor nt \rfloor \}$ and $I_r^{(n)}:=\{\lfloor nt \rfloor +1,\ldots, \lfloor nt \rfloor + \lfloor nh\rfloor\}$. We
set local estimators for $\mu^{\langle k \rangle }$ and $\mu^{\{ k \}}$ for $j\in \{\ell,r\}$ via
\begin{align}\label{moments}
\hat\mu_j^{\langle k \rangle} := 
\frac{1}{nh}
\sum\nolimits_{i\in I_j^{(n)}} X_i^k
\qquad\textrm{and}\qquad
\hat\mu_j^{\{k\}} :=
\frac{1}{nh}
\sum\nolimits_{i\in I_j^{(n)}} (X_i-\hat\mu_j^{\langle 1\rangle})^k,
\end{align}
and abbreviate $\hat{\mu}_j := \hat{\mu}_j^{\langle 1\rangle}$ and $\hat \sigma_j^{2} := \hat\mu_j^{\{2\}}$. Dependence on $t$ and $h$ is suppressed to avoid overload. Then we define
\begin{align}\label{dht}
D_{t,h}^{(n)} := \frac{\hat{\mu}_r - \hat{\mu}_\ell}{[(\hat \sigma_r^{2} + \hat \sigma_\ell^{2})/[nh]]^{1/2}},  
\end{align}
noting that $D_{t, h}^{(1)} = D_{t, h}$ from (\ref{Dintro}). We set $D_{t,h}^{(n)}:=0$ if the denominator vanishes. For each $n$ the statistics are c\`adl\`ag step-functions in both directions $t$ and $h$, with discontinuities in horizontal and vertical slices at values $k/n$, see \mbox{Figure \ref{fig:cadlag}}. 

Processes are considered in function space $(\mathcal D_{\rn} [\mathcal K],\|\cdot\|_{\infty})$. For a convex subset $\mathcal K$ of $\rn^k$ with $k\in\{1,2\}$ let $\mathcal D_{\rn} [\mathcal K]$ denote the set of $\rn$-valued functions on $\mathcal K$, which in case of $k=1$ are c\`adl\`ag, and in case of $k=2$ c\`adl\`ag with respect to each component when the other component is fixed. $\mathcal D_{\rn} [\mathcal K]$ is endowed with uniform distance $\|\cdot\|_{\infty}$. As asymptotics yield almost surely (a.s.) continuous limits, there is no need to evoke Skorokhod topology.

	\section{Asymptotics of the estimators}\label{sect:D_asymptotics}
\begin{figure}[h!]
	\centering  
	\includegraphics[width=0.65\textwidth,angle=0]{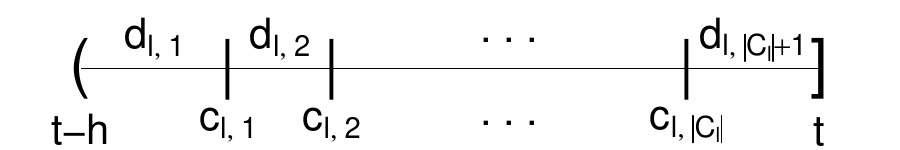}
	\vline\vline\vline\vline\vline\vline		
	\hspace{0.2em}\includegraphics[width=0.33\textwidth,angle=0]{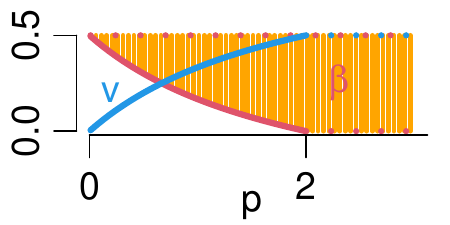}
	\caption{
		Left: schematic representation of the left window overlapping $|C_\ell|$ change points.
		Right: Parameters $v$ and $\beta$ depending on $p$.
	}
	\label{fig:leftwindow}
\end{figure}

\paragraph{Strong consistency}
We state strong convergence of the estimators. We define their limits pointwise for $(t,h)\in\Delta_\delta$. They are weightings of the theoretical moments depending on the change point locations relative to the window. Consider the left window $(t-h,t]$ and assume that the change points that are overlapped are $C_\ell\subset C$, i.e., we find $t-h< c_{\ell,1}<c_{\ell,2}< \cdots< c_{\ell,|C_\ell|}\le t$, and $c_{\ell, u}$ denotes the $u$-th smallest change point in $C_\ell$. Set $c_{\ell,0}:=t-h$ and $c_{\ell,|C_\ell|+1}:=t$. For $u=1,2,\ldots,|C_\ell|+1$ set the distance $d_{\ell,u}:=c_{\ell,u} - c_{\ell,u-1}$ between adjacent change points. The moments related to these sections are abbreviated by $\mu_{\ell,u}^{\langle k \rangle}$ and $\sigma_{\ell,u}^2$, see Figure \ref{fig:leftwindow} left.
From this we set
\begin{align}\label{tilde_left}
\tilde \mu_\ell^{\langle k\rangle } := 
\sum_{u=1}^{|C_\ell|+1} \frac{d_{\ell,u}}{h} \cdot \mu_{\ell,u}^{\langle k\rangle}
\qquad \textrm{and} \qquad
\tilde \sigma_\ell^2 := 
\sum_{u=1}^{|C_\ell|+1} \frac{d_{\ell,u}}{h} \cdot [\sigma_{\ell,u}^2 +(\tilde \mu_\ell -  \mu_{\ell,u})^2],
\end{align}
with $\tilde \mu_\ell := \tilde \mu_\ell^{\langle 1\rangle}$. The limits weight the moments with the time $d_{\ell,u}$ spent in a section relative to the window length $h=\sum_u d_{\ell,u}$. Further we set 
\begin{align}\label{error}
\tilde\tilde \sigma_\ell^2 := \sum_{u=1}^{|C_\ell|+1} \frac{d_{\ell,u}}{h} \cdot \sigma_{\ell,u}^2,
\qquad \textrm{and} \qquad
e_\ell^2:=\sum_{u=1}^{|C_\ell|+1} \frac{d_{\ell,u}}{h} (\tilde \mu_\ell -  \mu_{\ell,u})^2,
\end{align}
and note that $\tilde\sigma_\ell^2 = \tilde\tilde\sigma_\ell^2 + e_\ell^2$, i.e., $\tilde\sigma_\ell^2$ weights not only the variances $\sigma_{\ell,u}^2$ as in $\tilde\tilde \sigma_\ell^2$, but also the additional error term $e_\ell^2$. The error accounts for the violation of change points in the calculation of the mean in the computation of the empirical variance \--- the mean is calculated from all data in the windows 'violating' unknown change points. It is $e_\ell^2\ge 0$. If no change points are overlapped $C_\ell=\emptyset$, then the limits simplify to single theoretical moments $\tilde \mu_\ell^{\langle k\rangle }= \mu_{\ell,1}^{\langle k\rangle}$ and $\tilde\tilde\sigma_\ell^2=\tilde\sigma_\ell^2=\sigma_{\ell,1}^2$ and also $e_\ell^2=0$. 
Analogously we define the limits $\tilde \mu_r^{\langle k\rangle }$ and $\tilde \sigma_r^2$ for the right window $(t,t+h]$ by considering the change points $C_r$ whose ordered elements $c_{r,u}$ fulfill  $t< c_{r,1}<c_{r,2}< \cdots <c_{r,|C_r|}\le t+h$.
If $C=\emptyset$, then for all $(t,h)\in\Delta_\delta$ we obtain the population paramters $\tilde \mu_j=\mu$, $\tilde \sigma_j^2=\sigma^2$ etc.~ In the following we state strong consistency of the estimators.
\begin{lemm}\label{conv:est_cp}
	Let $\mathbf{X}\in\mathcal M$. For $j\in\{\ell,r\}$ it holds in $(\mathcal D_{\rn}[\Delta_\delta],\|\cdot\|_\infty)$ a.s.~as $n\to\infty$
	\begin{align*}
	n^v\cdot (\hat\mu_j^{\langle k\rangle} -\tilde\mu_j^{\langle k\rangle})_{(t,h)}&\longrightarrow (0)_{(t,h)}
	\; \textrm{for}\; v\in
	\begin{cases}
	(-\infty,1/2),\qquad\qquad  \qquad\,\,\,\,\hfill{\textrm{if}\,\, k=1}\\
	(-\infty,1/2)\cap(-\infty,p/(p+2)],\,\textrm{if}\, k=2,
	\end{cases}	
	\end{align*}	
	and $n^v\cdot (\hat\sigma_j^{2} -\tilde\sigma_j^{2})_{(t,h)}\to (0)_{(t,h)}$ with	$v\in(-\infty,1/2)\cap(-\infty,p/(p+2)]$.  
\end{lemm}	

The result follows from the Marcinkiewicz-Zygmund SLLN, noting that  $\ew[|X_i|^{2+p}]<\infty$. See $v$ in Figure \ref{fig:leftwindow} right: for $k=1$ we cannot reach $1/2$ noting CLT and LIL, and for $k=2$ the rate improves with the existence of higher moments via $p$, but only for $p<2$ while $p/(2+p)<1/2$, i.e., $p\ge 2$ brings no improvement. Any rate for $k=2$ is valid for $k=1$. 

The systematic component of $(D_{t,h}^{(n)})_{(t,h)\in \Delta_\delta}$ is the centering  $(d_{t,h}^{(n)})_{(t,h)\in\Delta_\delta}$ given by
\begin{align}\label{dth}
d_{t,h}^{(n)} := \frac{\tilde{\mu}_r - \tilde{\mu}_\ell}{[(\tilde \sigma_r^{2} + \tilde \sigma_\ell^{2})/[nh]]^{1/2}}.
\end{align}
We note that $n^{-1/2}\cdot d_{t,h}^{(n)}=d_{t,h}^{(1)}$ and conclude from Lemma \ref{conv:est_cp}

\begin{coro}\label{conv:Dth}
	Let $\mathbf{X}\in\mathcal M$. For $v\in(-\infty,1/2)\cap(-\infty,p/(p+2)]$ it holds in $(\mathcal D_{\rn}[\Delta_\delta],\|\cdot\|_\infty)$ a.s.~as $n\to\infty$
	\begin{align*}
	n^v \cdot 
	\Big(
	\frac{1}{\sqrt{n}} D_{t,h}^{(n)} - d_{t,h}^{(1)}
	\Big)_{(t,h)} \longrightarrow 
	(0)_{(t,h)}.			 
	\end{align*}		
\end{coro}	

\begin{figure}[h!]
	\centering  
	\includegraphics[width=0.8\textwidth,angle=0]{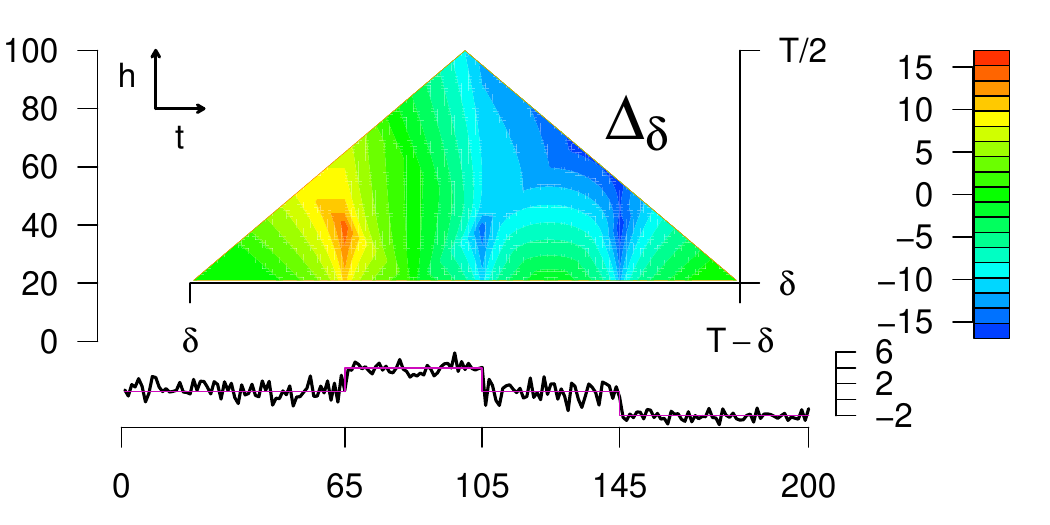}
	\caption{$d_{t,h}^{(1)}$ relating to $D_{t,h}^{(1)}$ from Figure \ref{fig:expl1a}.}
	\label{fig:expl1aa}
\end{figure}
Figure \ref{fig:expl1aa} shows $d_{t,h}^{(1)}$ relating to $D_{t,h}^{(1)}$. 
It has an analogous shape, but smoother as noise is canceled out. We will construct the zigzag-path from $D_{t,h}^{(n)}$ and deduce properties from a non-random path based on $d_{t,h}^{(n)}$.

\paragraph{Considering a single change point}
We describe $(d_{t,h}^{(n)})_t$ for fixed $h$ as a function of $t$.
Let $c_u\in C$. We assume $h$ small such that only $c_u$ is overlapped by the double-window when it is near $c_u$. We formalize the shape of $(d_{t,h}^{(n)})_{t}$ in Lemma \ref{lemm_shark}, which will be used to show that the zigzag-path is pushed towards $c_u$ when it comes close to it.

\begin{figure}[h!]
	\centering  
	\includegraphics[width=1\textwidth,angle=0]{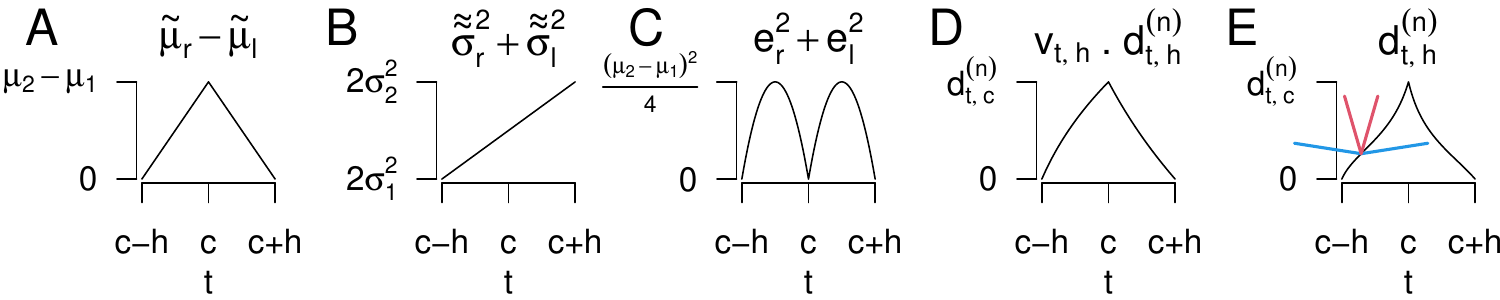}
	\caption{Construction of $(d_{t,h}^{(n)})_{t\in \tau_h}$ (for $c=c_1$). A: $\tilde \mu_r -\tilde \mu_\ell$, B: $\tilde\tilde \sigma_r^2+\tilde\tilde \sigma_\ell^2$, C: $e_r^2+e_\ell^2$,
		D: $v_{t,h}\cdot d_{t,h}^{(n)}$, E: $d_{t,h}^{(n)}$ and lines with slopes $\mp (nh)^{1/2}\cdot\kappa_a$ (blue) and $\mp (nh)^{1/2}\cdot\kappa_b$ (red).}
	\label{fig:shark}
\end{figure}

If $c_{u+1}-c_u \ge h$ and $c_{u}-c_{u-1} > h$,
then set $\tau_h := [c_{u-1}+h,c_{u+1}-h)$, else set $\tau_h:=\emptyset$. For $h$ small it is $\tau_h\not=\emptyset$. In this case $(d_{t,h}^{(n)})_{t\in \tau_h}$ is continuous, it vanishes if $|t-c_u|\ge h$ because no $c\in C$ is overlapped, and it systematically deviates from zero if $|t-c_u|< h$ with extreme value $t=c_u$, see Figure \ref{fig:shark}E	in which the blue and red lines indicate a uniform bound for the derivative. 
We set $v_{t,h}:=	[(\tilde\sigma_r^2 + \tilde\sigma_\ell^2)/(\tilde\tilde\sigma_r^2 + \tilde\tilde\sigma_\ell^2)]^{1/2}$,
recalling  $\tilde\sigma_j^2 = \tilde\tilde\sigma_j^2 + e_j^2$, see
(\ref{tilde_left}) and (\ref{error}), and denote $\wedge$ the minimum and $\vee$ the maximum.

\begin{lemm}\label{lemm_shark}
	Let $\mathbf X\in\mathcal M$, $c_u\in C$ and $h$ be fixed such that $\tau_h\not=\emptyset$. Then both $(d_{t,h}^{(n)})_{t\in\tau_h}$ and $(v_{t,h}\cdot d_{t,h}^{(n)})_{t\in\tau_h}$ are continuous and zero for $t \in\tau_h\backslash [c_u-h,c_u+h)$. 
	For $t \in [c_u-h,c_u+h)$, four cases are differentiated  regarding $d_{t,h}^{(n)}$: for $\mu_{u+1} > \mu_u$ we find that
	\begin{align}\label{dshape}
	d_{t,h}^{(n)} \quad\textrm{is}\;
	\begin{cases}
	\textrm{strictly increasing for t} \in [c_u-h,c_u],\\
	\textrm{strictly decreasing for t} \in (c_u,c_u+h].
	\end{cases}
	\end{align}
	For $\mu_{u+1} < \mu_u$ it is 'increasing' and 'decreasing' replaced. Further, for $t \in (c_u-h,c_u)\cup (c_u,c_u+h)$ it is	
	\begin{align}\label{derivative}
	\sqrt{nh}\cdot \kappa_a \le	
	\Big|\frac{\partial}{\partial t} d_{t,h}^{(n)}\Big|
	\le \sqrt{nh}\cdot \kappa_b,
	\end{align}
	with constants 
	\begin{align*}	
	\kappa_a &:=|\mu_{u+1}-\mu_{u}|\cdot\frac{2(\sigma_{u+1}^2 \wedge \sigma_u^2)}{[2(\sigma_{u+1}^2\vee \sigma_u^2)+(\mu_{u+1}-\mu_u)^2/4]^{3/2}},\\
	\kappa_b &:= |\mu_{u+1}-\mu_u|\cdot 
	\frac{2(\sigma_{u+1}^2\vee \sigma_u^2) + (\mu_{u+1}-\mu_u)^2}{[2(\sigma_{u+1}^2\wedge \sigma_u^2)]^{3/2}}.
	\end{align*}
	Regarding $v_{t,h}\cdot d_{t,h}^{(n)}$ six cases are differentiated:
	for $\mu_{u+1} > \mu_u$ and
	\begin{align*}
	\sigma_{u+1}^2= \sigma_u^2:&\quad
	v_{t,h}\cdot d_{t,h}^{(n)} \quad\textrm{is}\;
	\begin{cases}
	\textrm{linear and strictly increasing for t} \in [c_u-h,c_u],\\
	\textrm{linear and strictly decreasing for t} \in (c_u,c_u+h].
	\end{cases}\\
	\sigma_{u+1}^2> \sigma_u^2:&\quad 
	v_{t,h}\cdot d_{t,h}^{(n)} \quad\textrm{is}\;
	\begin{cases}
	\textrm{strictly concave and strictly increasing for t} \in [c_u-h,c_u],\\
	\textrm{strictly convex and strictly decreasing for t} \in (c_u,c_u+h].
	\end{cases}\\
	\sigma_{u+1}^2< \sigma_u^2:&\quad
	v_{t,h}\cdot d_{t,h}^{(n)} \quad\textrm{is}\;
	\begin{cases}
	\textrm{strictly convex and strictly increasing for t} \in [c_u-h,c_u],\\
	\textrm{strictly concave and strictly decreasing for t} \in (c_u,c_u+h].
	\end{cases}
	\end{align*}
	For $\mu_{u+1} < \mu_u$, the expressions hold true, but with 'convex' and 'concave' as well as 'increasing' and 'decreasing' replaced.  It is  $v_{t,h}\cdot |d_{t,h}^{(n)}|\ge |d_{t,h}^{(n)}|$ with equality at $t=c_u$.
\end{lemm}

Both $(v_{t,h}\cdot d_{t,h}^{(n)})_t$ and $(d_{t,h}^{(n)})_t$ are depicted in Figure \ref{fig:shark}D and E. 

Figure \ref{fig:shark} shows the statistics that factor into $(d_{t,h}^{(n)})_{t\in\tau_h}$. It is $\mu_2>\mu_1$ and $\sigma_2^2>\mu_2^2$. The function $\tilde \mu_r -\tilde \mu_\ell$ has the shape of a hat (A). It is positive as $\mu_2>\mu_1$. The function $\tilde\tilde \sigma_r^2+\tilde\tilde \sigma_\ell^2$ is linear (B). It is increasing as $\sigma_2^2>\sigma_1^2$. The error $e_r^2+e_\ell^2$ describes two parabolas and vanishes outside the $h$-neighborhood of $c:=c_u$ and also at $c$ (C). The error is maximal at $c\mp h/2$ which are the points where either the right or the left window is divided in half and thus equally sharing the left and right population. Panel D shows $v_{t,h}\cdot d_{t,h}^{(n)}$ which is $(nh)^{1/2}$ times (A) divided by the square-root of (B). It is non-negative as $\mu_2>\mu_1$ and concave on $[c-h,c]$ and convex on $(c,c+h]$ as $\sigma_2^2>\sigma_1^2$. Panel D shows $d_{t,h}^{(n)}$ which is $(nh)^{1/2}$ times (A) divided by the square-root of the sum of (B) and (C). As $\mu_2>\mu_1$, it is strictly increasing in $[c-h,c]$ and strictly decreasing on $(c,c+h]$. It is $v_{t,h}\cdot d_{t,h}^{(n)}\ge d_{t,h}^{(n)}$, as $v_{t,h}\ge 1$. Note that in Lemma \ref{lemm_shark} it is $0<\kappa_a<\kappa_b<\infty$, and $\kappa_a,\kappa_b$ depend only on the population parameters, such that the bounds for the derivative in (\ref{derivative}) depend on $h$ but not on $t$. The upper bound $(nh)^{1/2}\cdot\kappa_b$ is a Lipschitz constant for $(d_{t,h}^{(n)})_t$. Figure \ref{fig:shark}E shows lines with slopes $\mp (nh)^{1/2}\cdot\kappa_a$ (blue) and $\mp (nh)^{1/2}\cdot\kappa_b$ (red), i.e., the slope of $(d_{t,h}^{(n)})_t$ varies in between for all $t\in (c_u-h,c_u)\cup (c_u,c_u+h)$.	

\paragraph{Weak convergence} We state weak convergence of $(D_{t,h}^{(n)})_{(t,h)\in\Delta_\delta}$ if $C=\emptyset$.
Let $(W_t)_{t\ge0}$ be a standard Brownian motion. Then define $(L_{t,h})_{(t,h)\in\Delta_\delta}$ via
$L_{t,h}:=(2h)^{-1/2}\cdot[(W_{t+h} - W_t) - (W_t -W_{t-h})]$. 
The process $(L_{t,h})_{(t,h)}$ is continuous with $L_{t,h}\sim N(0,1)$. 

\begin{prop}\label{prop_main}
	Let $\mathbf X\in\mathcal M$ with $C=\emptyset$. In $(\mathcal D_{\rn}[\Delta_\delta],\|\cdot\|_\infty)$ it holds as $n\to\infty$ that $(D_{t,h}^{(n)})_{(t,h)}
	\stackrel{d}{\longrightarrow}
	(L_{t,h})_{(t,h)}$.
\end{prop}	

$L_{t,h}$ preserves the double-window structure of $D_{t,h}^{(n)}$, zero mean aligns with $C=\emptyset$ and unit variance results from scaling of $D_{t,h}^{(n)}$. The proof applies Donsker's theorom. 		

We mention that the upper limit theorems could be extended to dependent data structures, for which in  $D^{(n)}_{t, h}$ from (\ref{dht}) the variance estimator first needed to address that dependence. As a consequence, MSCP formulated in the following section would apply to those scenarios as well. For recent developments of change point detection methods w.r.t.~dependent data see e.g., \citep{Baranowski2019,Dette2020}, supplement \citep{Fryzlewicz2018s} of \citep{Fryzlewicz2018p}, or the overview article \citep{Aue2013}.
	
\section{Change point detection via MSCP}\label{sect:cpd}
\paragraph{Segmentation of the triangle}
We specify regions of $\Delta_\delta$. Each $c_u\in C$ has an area of attraction 
$A_u := \{(t,h)\in\Delta_\delta\,:\, c_{u-1} \le t-h < c_u\le t+h < c_{u+1} \}$, see Figure \ref{fig:attraction} (blue). For all $(t,h)\in A_u$ the associated double window overlaps $c_u$ but no other change point. It is $|d_{t,h}^{(n)}|>0$ for $(t,h)\in A_u$. We show that path that starts in $A_u$ is systematically pushed towards $c_u$, such that its endpoint becomes a proper estimate for $c_u$. We also set $B_u := \{(t,h)\in A_u \,:\, |t-c_u|\le h- \delta +1\}$, for whose elements we find the '$t$-distance' to $c_u$ smaller than the '$h$-distance' (plus $1$) to the bottom of $\Delta_\delta$. For any $t_0\in [\delta,T-\delta]$ we define its cone
$K_{t_0} := \{(t,h)\in \Delta_\delta \,:\, t-h<t_0\le t+h \}$. The cones $K_{c_u}$ of the $c_u\in C$ are shown in Figure \ref{fig:attraction} (orange). $K_{c_u}$ consists of all $(t,h)\in \Delta_\delta$ for which the double window overlaps $c_u$ and possibly neighboring change points. It holds $B_u \subset A_u\subset K_{c_u} \subset \Delta_\delta$.
Finally, we consider the remainder $R := \Delta_\delta \;\backslash\; \bigcup_{u=1}^{|C|} K_{c_u}$, that consists of $(t,h)\in\Delta_\delta$ for which no $c_u$ is overlapped by the double window, such that $d_{t,h}^{(n)}=0$ for all $(t,h)\in R$. In Figure \ref{fig:attraction}, $R$ consists of four subtriangles of $\Delta_\delta$ as there are three change points involved. 

\paragraph{The zigzag-downpath} We construct a path on $\Delta_\delta\cap \nn^2$ w.r.t.~either $D_{t,h}^{(n)}$ or $d_{t,h}^{(n)}$.
\begin{algo}\label{cons:zigzag} (zigzag-downpath)\\
	Input: $\mathbf{X}\in\mathcal M$ and a starting value $(t_s,h_s)\in\Delta_\delta\cap\nn^2$. Set $f_{t,h}^{(n)}\in \{d_{t,h}^{(n)},D_{t,h}^{(n)}\}$.\\	
	Construction: set 
	$t_s^{(n)}(0) :=\min\left(\argmax_{t\in \{t_s-1,t_s,t_s+1\}\cap \Delta_\delta} n^{-1/2}\cdot|f_{t,h_{s}}^{(n)}|\right).$
	For $k=1,2,\ldots,h_s- \delta$ iteratively define 
	\begin{align}\label{tsk}
	t_s^{(n)}(k) :=\min\left(\argmax_{t\in \{t_s^{(n)}(k-1)-1,t_s^{(n)}(k-1),t_s^{(n)}(k-1)+1\}} n^{-1/2}\cdot|f_{t,h_{s}-k}^{(n)}|\right).
	\end{align}
	Output: the zigzag downpath $(t_s^{(n)}(k),h_s-k)_{k=0,1,2,\ldots, h_s-\delta}$.	
\end{algo}
In $t_s^{(n)}(0)$ we start with choosing the maximizer among $\{t_s-1,t_s,t_s+1\}\cap \Delta_\delta$  and the restriction to $\Delta_\delta$ yields well-definedness if $(t_s,h_s)$ lies on its edge. The minimum in $t_s^{(n)}(k)$ ensures uniqueness. For $f_{t,h}^{(n)} = D_{t,h}^{(n)}$ the maximum is necessarily a.s.~unique in case the RVs of $\mathbf X$ lack point masses. 
We abbreviate the end $t_e^{(n)}:=t_s^{(n)}(h_s-\delta)$. In order to differentiate the paths we write 
\begin{align}
t_s^{(n)}(k)=:
\begin{cases}
t_s(k), &\textrm{if}\quad f_{t,h}^{(n)}=d_{t,h}^{(n)},\\
\hat t_s^{(n)}(k), &\textrm{if}\quad f_{t,h}^{(n)}=D_{t,h}^{(n)},\\	
\end{cases}
\end{align}	
and accordingly $t_e$ and $\hat t_e^{(n)}$, noting that for $f_{t,h}^{(n)}=d_{t,h}^{(n)}$ the path is independent of $n$, as $n^{-1/2}\cdot|d_{t,h}^{(n)}|=|d_{t,h}^{(1)}|$. 

\paragraph{Path behavior w.r.t~$d_{t,h}^{(n)}$} 
A path starting in $A_u$ systematically tends towards $c_u$:
\begin{lemm}\label{lemm:pathdth}
	Let $\mathbf{X}\in\mathcal M$, $c_u\in C$ and $f_{t,h}^{(n)}=d_{t,h}^{(n)}$.
	If $(t_s,h_s)\in A_u$, then for all $k\in\{1,2,\ldots, h_s- \delta\}$ it holds 
	\begin{align}\label{optimal:path}
	t_s(k)-t_s(k-1)=
	\begin{cases}
	0, &\textrm{if}\quad |t_s- c_u| \le k,\\
	1,&\textrm{if}\qquad t_s < c_u - k, \\
	-1,&\textrm{if}\qquad t_s > c_u + k,\\
	\end{cases}
	\end{align}
	If $(t_s,h_s)\in B_u$, then $t_e=c_u$. If $(t_s,h_s)\in A_u\backslash B_u$, then $|t_e-c_u|\le  \delta-1$. 
\end{lemm}

Lemma \ref{lemm:pathdth} states that in each step $t_s(k)$ either increases by unity if $t_s< c_u$ or it decreases by unity if $t_s> c_u$, and if it reaches $c_u$ then it stays. Thus, the path $(t_s(k),h_s-k)_{k}$ is perfectly zigzaging towards the vertical line at $c_u$. If it reaches this line then it moves vertically downwards to the lower edge of $\Delta_\delta$, i.e., $t_e=c_u$. The proof exploits the shape of $(d_{t,h}^{(n)})_t$ stated in Lemma \ref{lemm_shark}. 

\paragraph{Path behavior w.r.t~$D_{t,h}^{(n)}$} 
Regarding $D_{t,h}^{(n)}$, a path starting in $A_u$ converges towards the path w.r.t.~$d_{t,h}^{(n)}$:  

\begin{prop}\label{lemm:pathDth}
	Let $\mathbf X\in\mathcal M$, $c_u\in C$ and $f_{t,h}^{(n)}=D_{t,h}^{(n)}$. 
	Let $(t_s,h_s)\in A_u$, then for all $k\in\{0,1,2,\ldots, h_s- \delta\}$ it holds that
	$\hat t_s^{(n)}(k)\to t_s(k)$ a.s.~as $n\to\infty$ for all $k$, while $t_s(k)$ is given in (\ref{optimal:path}).
\end{prop}	
Consequently, starting in $A_u$ yields detection of $c_u$ up to a distance of $\delta -1$. More precisely, for $(t_s,h_s)\in B_u$ we find $\lim_{n\to\infty}\hat t_e^{(n)} = c_u$ a.s.,~and for $(t_s,h_s)\in A_u\backslash B_u$ it is $\lim_{n\to\infty}|\hat t_e^{(n)}-c_u| \le  \delta-1$ a.s.
The proof exploits proximity of $D_{t,h}^{(n)}$ and $d_{t,h}^{(n)}$.

\paragraph{Starting points} 
We call a set $S \subset  \nn^2 \cap \Delta_\delta$ a sufficient set of starting points if $S\cap A_u\not=\emptyset$ for all $u=1,\ldots,|C|$, i.e., for all $c_u \in C$ there lies a starting point in $A_u$.
For a mesh size $g\in \nn$ set a grid $S_g:= (g\cdot \nn^2)\cap \Delta_\delta$, see Figure \ref{fig:cutout}A (pink dots).

\begin{lemm}\label{lemm:scapa} Let $\mathbf X\in\mathcal M$ and $\delta < \lfloor \delta_C/2 \rfloor$. A grid $S_g$ is sufficient if $g\le \lfloor \delta_C/2 \rfloor$.
\end{lemm}	

The proof applies geometry. When a path enters $A_u$, then $c_u$ is detected up to a distance of $ \delta-1$, see Proposition \ref{lemm:pathDth}. Thus, when considering all paths starting in $S$, then all $c \in C$ will be detected. We need to avoid multiple detections of any $c$ as well as false positives. 

\paragraph{Change point detection} Set $\min \emptyset :=\infty$, and for a set $F$ denote $U(F)$ a uniformly sampled element.

\begin{algo}\label{algo:estimation} (MSCP)\\
	Input: $\mathbf X\in\mathcal M$, a set of starting values $S$,\\ 
	\phantom{AAA}and constants $\kappa>0$, and $\beta \in [1/2-v,1/2)$ for $v\in(0,1/2)\cap(0,p/(p+2)]$.\\	
	Initialize: counter $m=1$, estimates $\hat C_m^{(n)}=\emptyset$ and starting values
	$S_m^{(n)}=S$.\\ 
	While $S\not=\emptyset$, loop:\\
	1. Choose the starting value
	$(\hat t_{s,m}^{(n)},\hat h_{s,m}^{(n)}) := U(\argmax_{(t,h)\in S_{m}^{(n)}} [(nh)^{-1/2}\cdot|D_{t,h}^{(n)}|])$.\\
	2. Run the zigzag-path w.r.t.~$(\hat t_{s,m}^{(n)},\hat h_{s,m}^{(n)})$ and $f_{t,h}^{(n)}=D_{t,h}^{(n)}$,\\
	\phantom{AAA} call the path $(\hat t_{s,m}^{(n)}(k),\hat h_{s,m}^{(n)}-k)_{k=0,1,2,\ldots,\hat h_{s,m}^{(n)}- \delta}$ and the endpoint $(\hat t_{e,m}^{(n)}, \delta)$,\\
	\phantom{AAA} and compute the minimal distance $\mathscr D_m^{(n)}:=\min\{|\hat t_{e,m}^{(n)} - \hat c^{(n)}| : \hat c^{(n)} \in \hat C^{(n)}\}$.\\
	3. Check for multiple detections (3a.) and false positives (3b.)\\
	\phantom{AAA} 3a. If $\mathscr D_m^{(n)} \le 2(\delta-1)$, then set $S_{m}^{(n)} := S_{m}^{(n)} \backslash K_{\hat t_{e,m}^{(n)}}$ and Goto 1.\\ 
	\phantom{AAA} 3b. If $\max_{k=0,1\ldots,\hat h_{s,m}^{(n)}- \delta}$ $|D_{\hat t_{s,m}^{(n)}(k),\hat h_{s,m}^{(n)}-k}^{(n)}|<\kappa \cdot n^\beta$, then break the loop.\\
	\phantom{AAA} [3c. Optional: If $\mathscr D_m^{(n)} < \delta_C-2(\delta -1)$, then break the loop. (Needs input $\delta_C$.)]\\
	4. Update $m = m + 1$, then set $\hat C_m^{(n)} = \hat C_{m-1}^{(n)} \cup \{\hat t_{e,m-1}^{(n)}\}$ and $S_{m}^{(n)} = S_{m-1}^{(n)} \backslash K_{\hat t_{e,m-1}^{(n)}}$,\\
	\phantom{AAA}and Goto 1.\\ 
	Output: $\hat C^{(n)}:=\hat C_m^{(n)}$.
\end{algo}

The parameters $\kappa$ and $\beta$ yield a threshold $\kappa \cdot n^\beta$ (in 3b.), and if chosen large then this supports breaking the algorithm. Recall $E[|X_1|^{2+p}]<\infty$. $\beta$ is related to $p$ via $v$, see Figure \ref{fig:leftwindow} right. Small $p$ forces $\beta$ close to $1/2$ and a larger $p$ allows for smaller $\beta$. The parameter $\beta$ is redundant if $n=1$. In 1., the maximum is a.s.~unique if the RVs of $\mathbf X$ do not have point masses. In general, uniform sampling yields well-definedness. It is $\mathscr D_1^{(n)}=\infty$. 

Algorithm \ref{algo:estimation} (MSCP) is depicted in Figure \ref{fig:cutout}. Successively, $C$ is estimated. We comment on the steps.  
\begin{enumerate}
	\item[1.:] Among all current starting points the one maximizing $(nh)^{-1/2}\cdot|D_{t,h}^{(n)}|$ is chosen, uniformly in case of non-uniqueness. This ensures the starting point to lie outside the remainder $R$, a.s.~for $n$ large, in case of remaining undetected change points.
	\item[2.:] The zigzag-path is constructed and  $\hat t_{e,m}^{(n)}$ functions as a change point candidate. The minimal distance $\mathscr D_m^{(n)}$ of $\hat t_{e,m}^{(n)}$ and all previous estimates is computed. 
	\item[3.:] Two aspects are addressed: 
	\begin{enumerate}
		\item[3a.:] 'Avoiding multiple detections': if $\mathscr D_m^{(n)}\le2( \delta-1)$, then $\hat t_{e,m}^{(n)}$ is rejected, all starting points from its cone are cut, and estimation restarts. The rational is that if the paths of both $\hat t_{e,m}^{(n)}$ and the closest estimate passed $A_u$, then they also systematically led to $c_u$ up to $ \delta-1$. Hence, rejection of such $\hat t_{e,m}^{(n)}$ avoids multiple detections. 
		\item[3b.:] 'Avoiding false positives': if $\hat t_{e,m}^{(n)}$ was not rejected in 3a., then decision is made of whether the loop breaks. If the objection function takes extreme values on the zigzag-path, then this supports $\hat t_{e,m}^{(n)}$ to indicate an $c_u\in C$, and thus it is accepted in 4., and the next iteration starts after the cone of $\hat t_{e,m}^{(n)}$ is cut to avoid further detections of $c_u$.
	\end{enumerate}
\end{enumerate}
Note that we allow an error up to $ \delta-1$, and if the distance of $\hat t_{e,m}^{(n)}$ to $c_u$ is positive, then a slightly shifted cone was cut. This gives rise for 3a.~at first. Also note that if all $c\in C$ are detected and near cones are cut (again mentioning 3a.), then what is eventually left over is the remainder $R$. Thus, on the next path the objection function will no longer take extreme values which support breaking in 3b., and the previous estimates are returned. Finally note that the breaking criterion 3c.~applies if the minimal distance between change points was deceeded, also accounting for estimation precision only up to $ \delta -1$. Asymptotically, this will not occur, so 3c.~is obsolete for theoretical consistency, but may be used in practice.  

We denote $\hat c_u^{(n)}$ the $u$-th smallest element of $\hat C^{(n)}$. For well-definedness, if $|\hat C^{(n)}|< |C|$ set $\hat c_u^{(n)}:=0$ for all $u=|\hat C^{(n)}|+1,\ldots,|C|$.  The algorithm succeeds: 

\begin{theo}\label{theo:main}
	Let $\lfloor \delta_C/2 \rfloor > \delta$. In Algorithm \ref{algo:estimation} let the input $S$ be  sufficient. Then for the output $\hat C^{(n)}$ it holds a.s.~as $n\to\infty$ that 
	\begin{align*}
	|\hat C^{(n)}|\longrightarrow |C|\qquad \textrm{and}\qquad \limsup_{n\to\infty} |\hat c_u^{(n)} - c_u| \le \delta -1 \textrm{\; for all}\; u=1,\ldots,|C|.
	\end{align*}
\end{theo}

Location estimation is correct up to $ \delta  -1$. The choice $ \delta =1$ states strong consistency. Recall that $\lfloor \delta_C/2 \rfloor > \delta$ is needed for sufficiency of $S=S_g$ in Lemma \ref{lemm:scapa}. Also it implies $\delta_C > 2(\delta-1)$, i.e., neighboring change points should be separated by more than the worst errors of their estimators. The proof applies Proposition \ref{lemm:pathDth} for the estimation of $C$, and Corollary \ref{conv:Dth} for correct breaking.

\section{Practical aspects}\label{sect:practical}
\paragraph{Parameter choice} 	
We give recommendations for the choice of $S$, $\kappa$ and $\delta$ in practice, where $n=1$. Then, the threshold is $\kappa \cdot n^{\beta}=\kappa$, i.e., $\beta$ is redundant. 

\begin{figure}[h!]
	\centering  
	\includegraphics[width=0.3\textwidth,angle=0]{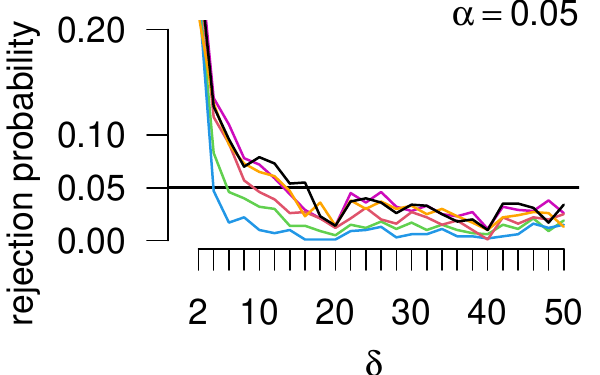}
	\hspace{0em}
	\includegraphics[width=0.3\textwidth,angle=0]{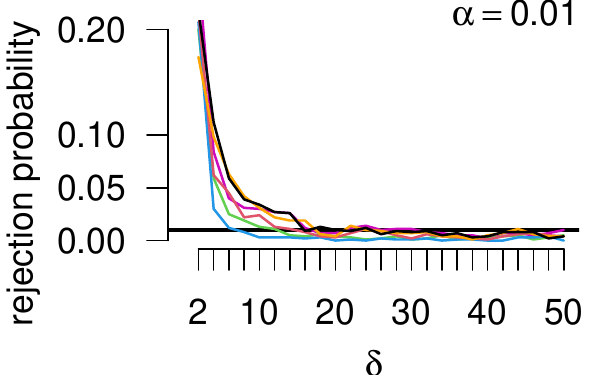}
	\includegraphics[width=0.3\textwidth,angle=0]{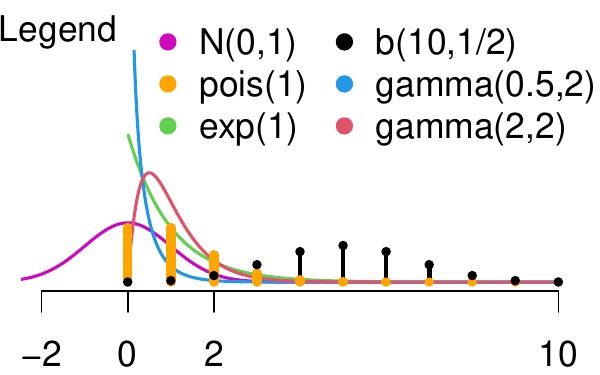}
	\caption{Rejection probability under $C=\emptyset$ ($1000$ simulations) depending on $\delta\in\{2,4,\ldots,50\}$. $T=1000$, left: $\alpha =5\%$, right: $\alpha =1\%$. Six distributions color coded: $N(0,1)$ (magenta), $Pois(1)$ (orange), $exp(1)$ (green), $b(10,1/2)$ (black), $gamma(0.5,2)$ (blue) and $gamma(2,2)$ (red). Right: Legend.}
	\label{fig:rej}
\end{figure}

\begin{enumerate}
	\item[1.] 
	Choice of $\kappa$: $\kappa$ can be chosen as the rejection threshold of a test for $H_0:C=\emptyset$: For $H_0$ Proposition \ref{prop_main} implies weak convergence of $\sup_{t,h\in\Delta_\delta}|D_{t,h}^{(n)}|\to\sup_{(t,h)\in\Delta_\delta}|L_{(t,h)}|$ as $n\to\infty$, and we choose $\kappa$ as the $(1-\alpha)$-quantile of the limit distribution (e.g., significance level $\alpha=0.01$), which can be derived in simulations, compare e.g., \citet{Messer2014}. If $|\hat C^{(1)}|>0$, then $H_0$ is rejected at level $\le\alpha$, as the maximum w.r.t.~a path is bounded by the supreme over $\Delta_\delta$. 
	
	\item[2.] Choice of $S$: Lemma \ref{lemm:scapa} states sufficiency of $S=S_g$ if $g\le \lfloor \delta_C/2\rfloor$ (in case $\delta < \lfloor \delta_C/2\rfloor$). Thus, if $\delta_C$ is known it is reasonable to choose $g=\lfloor\delta_C/2\rfloor$ as $g$ large reduces computational complexity. If $\delta_C$ is unknown, we set $g=\delta$ or even smaller if computational complexity allows for.    
	
	\item[3.] Choice of $\delta$: We recommend $\delta=20$: by construction, asymptotic considerations imply an increase of the double windows. For $n\to\infty$, weakly $D_{t,h}^{(n)}\to L_{t,h}\sim N(0,1)$ for all $(t,h)\in\Delta_\delta$, but for $n=1$ small $h$ means less observations and the approximation via $N(0,1)$ is harder to justify. Practically, for tiny $h$ there is high variability in $\hat \mu_j$ and $\hat \sigma_j^2$ and thus $D_{t,h}^{(1)}$ is likely to take extreme values even if no change is involved, resulting in false positives. Thus, $\delta $ should be bounded from below.
	Figure \ref{fig:rej} shows simulations for the probability that $\pr(|\hat C^{(1)}|>0)$ if $C=\emptyset$ (rejection probability under $H_0$), for both $\alpha=0.05$ and $\alpha=0.01$, depending on $\delta$. Over a variety of distributions, including normal, exponential, gamma, binomial and Poisson, for $\delta\ge20$ the $\alpha$-level is kept throughout.
\end{enumerate}

In the following we use $\delta=g=20$ and $\kappa$ is derived in simulations for $\alpha=0.01$.

\paragraph{Simulation studies}
We evaluate the performance of MSCP in simulation studies. Throughout it is $T=1000$ and $|C|=5$. We consider different scenarios of locations $c_u\in C$ and parameters $\mu_u$ and $\sigma_u$, given in Table \ref{tab:scenario}. 
Scenarios 1\---3 differ in $C$: in 1 it is $c_{u+1}-c_u = 200$, and in 2 it is $c_{u+1}-c_u = 100$ throughout, and 3 describes a multiscale setup where $c_1=200$ is well separated and for the others it is $c_{u+1}-c_u = 50$.
For each choice of $C$ we then consider different effects by altering both changes in $\mu_u$ and in $\sigma_u$. E.g., 1a states certain $\mu_u$ which are halved in 1c. Also, in 1a it is $\sigma_u=1$ constant, while in 1b the $\sigma_u$ alter between $1$ and $2$. 
For each scenario we consider different distributions: A.~$N(\mu,\sigma^2)$ (normal), B.~$gamma(s,\lambda)$ (gamma), C.~$Poi(\lambda)$ (Poisson), D.~$b(10,p)$ (binomial, $n=10$ fix), and E.~is a combination of the previous that changes between the six sections using A.,B.,C.,D.,A.,B. (mix). Note that for $Poi(\lambda)$ and $b(10,p)$ the formulation of $\sigma_u$ can be disregarded as it follows from $\mu_u$, and also if distributions coincide due to equality of $\mu_u$ as e.g., in 1a and 1b, they are presented only once.

\begin{table}[h!]
	\begin{tabular}{| l | l | l | l |}
		\hline
		Scenario & $C$ & $\mu_u$ & $\sigma_u$ \\
		\hline	 \hline
		1a & $100,300,500,700,900$ & $1,4,1,8,1,4$ & $1,1,1,1,1,1$\\ 
		\hline	 
		1b & $100,300,500,700,900$ & $1,4,1,8,1,4$ & $1,2,1,2,1,2$\\ 
		\hline	 
		1c & $100,300,500,700,900$ & $0.5,2,0.5,4,0.5,2$ & $1,1,1,1,1,1$\\ 
		\hline\hline
		2a & $300,400,500,600,700$ & $1,4,1,8,1,4$ & $1,1,1,1,1,1$\\ 
		\hline	 
		2b & $300,400,500,600,700$ & $1,4,1,8,1,4$ & $1,2,1,2,1,2$\\ 
		\hline	 
		2c & $300,400,500,600,700$ & $0.5,2,0.5,4,0.5,2$ & $1,1,1,1,1,1$\\ 
		\hline\hline	 
		3a & $200,500,550,600,750$ & $1,4,1,8,1,4$ & $1,1,1,1,1,1$\\
		\hline	 
		3b & $200,500,550,600,750$ & $1,4,1,8,1,4$ & $1,2,1,2,1,2$\\
		\hline	 
		3c & $200,500,550,600,750$ & $0.5,2,0.5,4,0.5,2$ & $1,1,1,1,1,1$\\
		\hline	 
		3d & $200,500,550,600,750$ & $0.5,2,0.5,4,0.5,2$ & $1,2,1,2,1,2$\\
		\hline
		3e & $200,500,550,600,750$ & $1,2,4,8,4,2$ & $1,1,1,1,1,1$\\
		\hline	 
		\hline 
	\end{tabular} 
	
	\vspace{0.2cm}
	\noindent\caption{Scenarios of $C$, $\mu$ and $\sigma$ considered in simulations.}
	\label{tab:scenario}
\end{table}

For each scenario we run $1000$ simulations, i.e., there are $1000 \cdot |C|=5000$ change points in total. W.r.t.~a critical value $\mathcal V$ we classify an estimate $\hat c$ as correct if $m_{\hat c} :=\min_{c\in C} |\hat c - c|\le \mathcal V$. Let $\hat C_{\mathcal V}$ denote the set of all correct estimates w.r.t.~$\mathcal V$. Then $\hat C_T$ is the set of all estimates. We compute $|\hat C_T|$, and for $\mathcal V\in \{10,5,2\}$ both $|\hat C_{\mathcal V}|$ and the mean absolute deviation $M_{\mathcal V}:=|\hat C_{\mathcal V}|^{-1} \sum_{\hat c\in \hat C_{\mathcal V}} m_{\hat c}$. 
The results, depending on the scenarios and  distributions, are reported in \mbox{Tables \ref{tab:results1}\---\ref{tab:results3}}. 

\begin{table}[h!]
	\begin{tabular}{| l | l || l | l | l | l |}
		\hline
		Scenario & Distribution & $|\hat C_{T}|$ & $|\hat C_{ 10}|$, $M_{10}$  & $|\hat C_{5}|$, $M_{5}$ & $|\hat C_{ 2}|$, $M_{2}$\\
		\hline	 \hline
		1a & A.~(normal)   &5005 &5000, 0.1& 5000, 0.1& 4994, 0.1\\
		\hline
		1a & B.~(gamma)    &5002 &5000, 0.1& 5000, 0.1& 4993, 0.1\\
		\hline
		1a & C.~(Poisson)     &5005 &5000, 0.4& 4993, 0.3& 4905, 0.3\\
		\hline
		1a & D.~(binomial) &5005 &5000, 0.2& 4998, 0.2& 4954, 0.2\\
		\hline
		1a & E.~(mix)      &5001 &5000, 0.1& 5000, 0.1& 4993, 0.1\\
		\hline 
		\hline
		1b & A.~(normal)   &5001 &4998, 0.3& 4993, 0.3& 4906, 0.2\\
		\hline
		1b & B.~(gamma)    &5000 &5000, 0.3& 4997, 0.3& 4909, 0.2\\
		\hline
		1b & E.~(mix)      &5004 &5000, 0.3& 4999, 0.3& 4906, 0.2\\
		\hline 
		\hline
		1c & A.~(normal)   &4951 &4935, 0.5& 4912, 0.5& 4698, 0.4\\
		\hline
		1c & B.~(gamma)    &4953 &4932, 0.4& 4925, 0.4& 4823, 0.3\\
		\hline
		1c & C.~(Poisson)     &4640 &4626, 0.6& 4600, 0.6& 4370, 0.5\\
		\hline
		1c & D.~(binomial) &4891 &4883, 0.6& 4858, 0.5& 4642, 0.4\\
		\hline
		1c & E.~(mix)      &4936 &4929, 0.6& 4903, 0.5& 4707, 0.4\\
		\hline	 
		\hline
	\end{tabular}
	\vspace{0.2cm}
	
	\caption{Simulation results for scenarios 1a\---c.}
	\label{tab:results1}
\end{table}

\begin{table}[h!]
	\begin{tabular}{| l | l || l | l | l | l |}
		\hline
		Scenario & Distribution & $|\hat C_{T}|$ & $|\hat C_{ 10}|$, $M_{10}$  & $|\hat C_{5}|$, $M_{5}$ & $|\hat C_{ 2}|$, $M_{2}$\\
		\hline	 \hline
		2a & A.~(normal)   &5002 &5000, 0.1 &5000, 0.1& 4993, 0.1\\
		\hline
		2a & B.~(gamma)    &5002 &5000, 0.1 &5000, 0.1& 4988, 0.1\\
		\hline
		2a & C.~(Poisson)     &5005 &5000, 0.3 &4998, 0.3& 4908, 0.3\\
		\hline
		2a & D.~(binomial) &5002 &5000, 0.2 &5000, 0.2& 4958, 0.2\\
		\hline
		2a & E.~(mix)      &5002 &5000, 0.1 &5000, 0.1& 4989, 0.1\\
		\hline	 
		\hline
		2b & A.~(normal)   &5002& 4998, 0.3& 4995, 0.3& 4914, 0.2\\
		\hline
		2b & B.~(gamma)    &5003& 5000, 0.3& 4998, 0.3& 4928, 0.2\\
		\hline
		2b & E.~(mix)      &5008& 5000, 0.3& 4998, 0.3& 4935, 0.2\\
		\hline 
		\hline
		2c & A.~(normal) &4884 &4873, 0.5 &4855, 0.5 &4663, 0.4\\
		\hline
		2c & B.~(gamma) &4906 &4870, 0.4 &4863, 0.3 &4766, 0.3\\
		\hline
		2c & C.~(Poisson) & 4553 &4541, 0.7 &4520, 0.6 &4285, 0.5\\
		\hline
		2c & D.~(binomial) & 4847 &4841, 0.5 &4828, 0.5 &4647, 0.4\\
		\hline
		2c & E.~(mix) &4926 &4920, 0.5 &4902, 0.5 &4720, 0.4\\
		\hline 
	\end{tabular}
	\vspace{0.2cm}
	
	\caption{Simulation results for scenarios 2a-c.}
	\label{tab:results2}
\end{table}

\begin{table}[h!]
	\begin{tabular}{| l | l || l | l | l | l |}
		\hline
		Scenario & Distribution & $|\hat C_{T}|$ & $|\hat C_{ 10}|$, $M_{10}$  & $|\hat C_{5}|$, $M_{5}$ & $|\hat C_{ 2}|$, $M_{2}$\\
		\hline
		3a & A.~(normal)   &5004& 4990, 0.3& 4910, 0.2& 4846, 0.1\\
		\hline
		3a & B.~(gamma)    &5002& 4990, 0.3& 4879, 0.2& 4828, 0.1\\
		\hline
		3a & C.~(Poisson)     &4942& 4876, 0.9& 4580, 0.5& 4334, 0.3\\
		\hline
		3a & D.~(binomial) &5005& 4965, 0.6& 4787, 0.3& 4652, 0.2\\
		\hline
		3a & E.~(mix)      &5000& 4991, 0.3& 4901, 0.1& 4857, 0.1\\
		\hline 
		\hline
		3b & A.~(normal)   &4988& 4928, 0.8& 4657, 0.4& 4478, 0.3\\
		\hline
		3b & B.~(gamma)    &4995& 4939, 0.8& 4689, 0.4& 4482, 0.2\\
		\hline
		3b & E.~(mix)      &5001& 4947, 0.6& 4789, 0.3& 4638, 0.2\\
		\hline	 
		\hline
		3c & A.~(normal)   &4814& 4703, 1.3& 4286, 0.7& 3936, 0.4\\
		\hline
		3c & B.~(gamma)    &4820& 4749, 1.2& 4334, 0.5& 4095, 0.3\\
		\hline
		3c & C.~(Poisson)     &4387& 4249, 1.5& 3845, 0.8& 3480, 0.5\\
		\hline
		3c & D.~(binomial) &4750& 4644, 1.5& 4146, 0.7& 3809, 0.4\\
		\hline
		3c & E.~(mix)      &4856& 4756, 1.3& 4334, 0.7& 3990, 0.4\\
		\hline	 
		\hline
		3d & A.~(normal)   &3093& 2946, 1.6& 2620, 0.8& 2380, 0.5\\
		\hline
		3d & B.~(gamma)    &2962& 2842, 1.8& 2472, 0.8& 2193, 0.4\\
		\hline
		3d & E.~(mix)      &2896& 2804, 1.4& 2548, 0.7& 2296, 0.4\\
		\hline 
		\hline
		3e & A.~(normal)   &4680& 4566, 1.1& 4257, 0.5& 3984, 0.3\\
		\hline
		3e & B.~(gamma)    &4735& 4620, 1.0& 4302, 0.5& 4046, 0.3\\
		\hline
		3e & C.~(Poisson)     &3411& 3085, 2.3& 2584, 1.2& 2154, 0.6\\
		\hline
		3e & D.~(binomial) &4129& 3883, 1.8& 3346, 0.8& 2959, 0.4\\
		\hline
		3e & E.~(mix)      &4220& 4014, 1.8& 3452, 0.8& 3070, 0.4\\
		\hline	 
	\end{tabular}
	\vspace{0.2cm}
	\caption{Simulation results for scenarios 3a-e.}
	\label{tab:results3}
\end{table}

We summarize the results. Scenarios 1a and 1b are well-posed  in the sense that results are strong and yield precise estimates: $|\hat C_T|$ is close to $5000$ over all scenarios and distributions. Notably, in 1a it is exactly $|\hat C_{10}|=5000$ throughout, and $|\hat C_5|=5000$ for three distributions. Also, the strict measure $|\hat C_2|$ reveals precise results. In 1b we consider varying $\sigma_u$, but the estimates remain similarly precise. The setups 2a and 2b are again well-posed: 2a and 2b are pretty similar to 1a and 1b. In setups 3a and 3b performance is slightly reduced as compared to 2a and 2b due to closer distances. Nevertheless, we obtain about $5000$ estimates at precise location for 3a and 3b. In scenarios 1c, 2c and 3c we have jump sizes halved as compared to 1a-3a and 1b-3b. Here, deficiencies start to arise: $|\hat C_T|$ reduces to about $4750$ to $4950$ for distributions A., B., D. and E., while for C., it drops to about $4400$ to $4800$. In scenario 3d we reduced the effects w.r.t.~both smaller jump sizes and higher variances, resulting in higher performance losses as we obtain only around $3000$ estimates. But note that e.g., for $c_6$ it is $\mu_6-\mu_5=1.5$ with $\sigma_6=2$ which is reasonably hard to be detected, also taking into account the smaller spacings between $c_u$. Despite the reduced number, the  precision for detected estimates is kept high. Scenario 3e is another example with different jump sizes resulting in a good overall outcome, but slightly weaker than 3a as effects are reduced similarly to 3c. Overall, MSCP yields reliable results over different scenarios of change points and effects: both the number and the locations are reliably estimated. Notably, performance is also kept over different distributions and regardless of changes in variance.

\paragraph{Comparison with state of the art methods}
We present simulation studies which reveal competitiveness of \texttt{mscp} with existing change point detection methods available on \texttt{CRAN}. For all competitors we consider the scenarios 1c, 2c and 3c, see Table \ref{tab:scenario}, and all distributions A.\---E.  from the previous paragraph. The reason for the choice of theses scenarios is that for MSCP deficiencies in detection started to arise. Again, for each such setup we run $1000$ simulations which yields $5000$ change points in total. 

Note that between the three scenarios the locations of $C$ vary while $\mu_u=0.5,2,0.5,4,0.5,2$ and $\sigma_u = 1,\ldots,1$ remain unchanged. In particular, due to constant $\sigma_u$, results should not be affected through a method's capability of handling changes in variance, which holds true for MSCP. For that also keep in mind that, in fact, $\sigma_u\equiv 1$ only holds true for A.~B.~and parts of E., while for C.~and~D.~$\sigma_u$ always changes with $\mu_u$.     

As competitors of \texttt{mscp} we consider \texttt{mosum} \citet{CRANmosum} with individual windows \texttt{G=50}, \texttt{100} and \texttt{200}, respectively, \texttt{wbs} \citet{CRANwbs}, \texttt{not} \citet{CRANnot} with \texttt{contrast=}\texttt{pcwsConst}\-\texttt{MeanVar} and \texttt{pcwsConstMeanHT}, respectively, 
\texttt{changepoint} \citet{CRANchangepoint} considering \texttt{cpt.meanvar} with \texttt{method=}\texttt{PELT} and \texttt{test.stat=}'\texttt{Normal}' for A. D. and E., '\texttt{Gamma}' for B., and '\texttt{Poisson}' for C., 
\texttt{stepR} \citet{CRANstepR} with \texttt{stepFit}, \texttt{cumSeg} \citet{CRANcumSeg} using \texttt{jumppoints}, and \texttt{FDRSeg} \citet{CRANFDRSeg} using \texttt{fdrseg} with \texttt{sd} estimating the global standard deviation. Any other tuning-parameters are kept default. 

The results are presented in Table \ref{tab:comparison_1a} (scenario 1c), Table \ref{tab:comparison_2a} (scn.~2c) and \mbox{Table \ref{tab:comparison_3a}} (scn.~3c). We first recall that overall, \texttt{mscp} yields a total number of estimates  $|\hat C_T|$ of about $4750$ to $4950$ for distributions A., B., D.~and E., and about $4400$ to $4800$ for C. Further, recall high estimation precision regarding different levels of tolerance, i.e., values of $|\hat C_{\mathcal V}|$ and $M_{\mathcal V}$ for $\mathcal V=\{10,5,2\}$. In total, also the competitors often show a similar performance in many scenarios, but also somewhat weaken in certain setups. For \texttt{mosum} we nicely see, e.g., in Table \ref{tab:comparison_1a} A., that performance depends on the selection of the individual window. For example we find \texttt{G=50} to slightly overestimate, and \texttt{G=200} to underestimate the total number of change points, while the middle window \texttt{G=100} performs best, also noting slighty reduced location precision for small windows as compared to $\texttt{mscp}$. Note that overall the problem of a fixed window becomes even more evident when change point distances are reduced, see Table \ref{tab:comparison_2a} and \ref{tab:comparison_3a}, particularly because large windows lose sensitivity. Of course, window selection is a major challenge in practice as change point locations are unknown, and $\texttt{mscp}$ circumvents it by nature. Apart form this, \texttt{mosum} yields quite stable results among different distributions, but similar to \texttt{mscp} it loses power for distribution C.
For \texttt{wbs} we find good performance for distribution A., w.r.t.~both the number and precision of estimates over all scenarios, while it tends to overestimate for other distributions, most extreme for C.~and B. We overall find good performance for the method \texttt{not}. However, we mention that \texttt{contrast=}\texttt{pcwsConstMeanVar} tends to overestimate particularly for distribution B., noting that the contrast assumes normality. Interestingly, \texttt{contrast=}\texttt{pcwsConstMeanHT}, which assumes heavy-tailed additive errors, overall performs nicely, while showing somewhat reduced estimation precision, as compared to \texttt{mscp}. For \texttt{changepoint} we find great performance for A.~and C.~where parametric assumptions are met, but tremendous overestimation if not, see D.  and E. Although parametric assumptions are met in B.,~it also overestimates. 
\texttt{stepR} shows reliable results in A., while it tends to overestimate the number of change points if normality is violated. This tendency is less drastic in E.~where the distribution is normal in at least two sections, or D., but more problematic for distribution B.
Over all setups \texttt{cumSeg} perform nicely and only very slightly overestimates the number of change points, but shows somewhat reduced estimation precision as compared to \texttt{mscp}. 
\texttt{FDRSeg} also shows good performance among almost all setups presented, w.r.t.~both different scenarios and distributions. However, we mention overestimation in C., and a slight reduction in estimation precision as compared to \texttt{mscp}.

In summary, the results presented reveal serious competitiveness of \texttt{mscp} with state of the art change point detection methods. Particularly, \texttt{mscp} overcomes the selection of a tuning bandwidth as compared to single window MOSUM techniques. Also, the consideration of scenarios 1c\---3c reveals comparably good performance regarding different change point distances. Besides that, a comparably stable performance over different distributions supports the value of the non-parametric nature of \texttt{mscp} for practice.

\begin{table}[h!]
	\begin{scriptsize}
		\begin{tabular}{| l | l | l || l | l | l | l |}
			\hline
			Scn. & Distr. & Method & $|\hat C_{T}|$ & $|\hat C_{ 10}|$, $M_{10}$  & $|\hat C_{5}|$, $M_{5}$ & $|\hat C_{ 2}|$, $M_{2}$\\
			\hline	 \hline
			1c & A.~(normal) &  \texttt{mscp}  &4951 &4935, 0.5& 4912, 0.5& 4698, 0.4\\ \hline
			& &  \texttt{mosum} $G=50$   			 & 5057 & 4965 , 0.9 & 4809 , 0.7 & 4378 , 0.4 \\				\hline
			& &  \texttt{mosum} $G=100$	  		 & 5002 & 4969 , 0.8 & 4852 , 0.7 & 4430 , 0.4 \\				\hline
			& &  \texttt{mosum} $G=200$	   		 & 4002 & 3972 , 0.4 & 3905 , 0.3 & 3757 , 0.2 \\				\hline
			& &  \texttt{wbs}   			 & 5030 & 4983 , 0.7 & 4890 , 0.6 & 4529 , 0.4 \\				\hline
			& &  \texttt{not} '\texttt{MeanVar}'  			 & 5007 & 4986 , 0.7 & 4886 , 0.6 & 4517 , 0.4 \\	\hline
			& &  \texttt{not} '\texttt{MeanHT}'  			 & 5010 & 4974 , 0.8 & 4846 , 0.6 & 4438 , 0.4 \\	\hline
			& &  \texttt{changepoint}	   		 & 5033 & 5002 , 0.8 & 4893 , 0.6 & 4523 , 0.4 \\				\hline
			& &  \texttt{stepR}	  		 & 5037 & 4978 , 1.4 & 4863 , 1.2 & 4443 , 1 \\				\hline
			& &  \texttt{cumSeg}   			 & 5019 & 4935 , 1.5 & 4782 , 1.4 & 3953 , 0.9 \\				\hline
			& &  \texttt{FDRSeg} 			 & 5000 & 4979 , 1.4 & 4867 , 1.2 & 4446 , 1 \\				\hline			
			\hline
			& B.~(gamma) &\texttt{mscp}    &4953 &4932, 0.4& 4925, 0.4& 4823, 0.3\\ \hline
			& &  \texttt{mosum} $G=50$   			 & 5082 & 4873 , 1 & 4658 , 0.7 & 4223 , 0.4 \\				\hline
			& &  \texttt{mosum} $G=100$	  		 & 5014 & 4889 , 0.9 & 4696 , 0.6 & 4286 , 0.4 \\				\hline
			& &  \texttt{mosum} $G=200$	   		 & 4005 & 3948 , 0.4 & 3881 , 0.3 & 3748 , 0.2 \\				\hline
			& &  \texttt{wbs}   			 & 29906 & 6827 , 2.1 & 5692 , 0.9 & 4846 , 0.4 \\				\hline
			& &  \texttt{not} '\texttt{MeanVar}'  			 & 15799 & 5076 , 1.9 & 4342 , 0.8 & 3738 , 0.3 \\			\hline
			& &  \texttt{not} '\texttt{MeanHT}'  			 & 5029 & 4991 , 0.6 & 4928 , 0.5 & 4671 , 0.3 \\			\hline
			& &  \texttt{changepoint}	   		 & 14342 & 5529 , 2.6 & 4397 , 1.3 & 3390 , 0.5 \\				\hline
			& &  \texttt{stepR}	  		 & 30957 & 6568 , 2.7 & 5408 , 1.5 & 4504 , 1.1 \\				\hline
			& &  \texttt{cumSeg}   			 & 5033 & 4911 , 1.5 & 4734 , 1.3 & 4015 , 0.9 \\				\hline
			& &  \texttt{FDRSeg} 			 & 6230 & 4922 , 1.5 & 4760 , 1.3 & 4330 , 1 \\				\hline
			\hline
			& C.~(Poisson)  & \texttt{mscp}   &4640 &4626, 0.6& 4600, 0.6& 4370, 0.5\\ \hline		
			& &  \texttt{mosum} $G=50$   			 & 4812 & 4679 , 1.1 & 4512 , 0.9 & 4002 , 0.5 \\				\hline
			& &  \texttt{mosum} $G=100$	  		 & 4847 & 4752 , 1.1 & 4561 , 0.8 & 4048 , 0.5 \\				\hline
			& &  \texttt{mosum} $G=200$	   		 & 3900 & 3886 , 0.5 & 3844 , 0.4 & 3672 , 0.2 \\				\hline
			& &  \texttt{wbs}   			 & 13155 & 5447 , 1.6 & 4948 , 0.9 & 4276 , 0.5 \\				\hline
			& &  \texttt{not} '\texttt{MeanVar}'  			 & 5231 & 4941 , 1 & 4755 , 0.8 & 4276 , 0.4 \\				\hline
			& &  \texttt{not} '\texttt{MeanHT}'  			 & 5090 & 4906 , 1.1 & 4679 , 0.8 & 4182 , 0.5 \\			\hline
			& &  \texttt{changepoint}	   		 & 5000 & 4964 , 0.9 & 4834 , 0.8 & 4362 , 0.4 \\				\hline
			& &  \texttt{stepR}	  		 & 12518 & 5253 , 2.1 & 4743 , 1.5 & 3953 , 1.1 \\				\hline
			& &  \texttt{cumSeg}   			 & 5021 & 4896 , 1.8 & 4637 , 1.5 & 3700 , 1 \\				\hline
			& &  \texttt{FDRSeg} 			 & 5253 & 4921 , 1.7 & 4685 , 1.4 & 4030 , 1.1 \\				\hline
			\hline
			& D.~(binomial) & \texttt{mscp} &4891 &4883, 0.6& 4858, 0.5& 4642, 0.4\\ \hline
			& &  \texttt{mosum} $G=50$   			 & 4837 & 4739 , 0.9 & 4585 , 0.7 & 4165 , 0.4 \\				\hline
			& &  \texttt{mosum} $G=100$	  		 & 4846 & 4791 , 1 & 4622 , 0.7 & 4189 , 0.4 \\				\hline
			& &  \texttt{mosum} $G=200$	   		 & 3911 & 3896 , 0.4 & 3854 , 0.3 & 3715 , 0.2 \\				\hline
			& &  \texttt{wbs}   			 & 6472 & 5035 , 1 & 4816 , 0.7 & 4415 , 0.4 \\				\hline
			& &  \texttt{not} '\texttt{MeanVar}'  			 & 5184 & 4972 , 0.9 & 4798 , 0.7 & 4399 , 0.4 \\			\hline
			& &  \texttt{not} '\texttt{MeanHT}'  			 & 5055 & 4952 , 1 & 4764 , 0.7 & 4334 , 0.4 \\				\hline
			& &  \texttt{changepoint}	   		 & 191707 & 21855 , 4.9 & 12235 , 2.5 & 6215 , 1 \\				\hline
			& &  \texttt{stepR}	  		 & 6369 & 4987 , 1.6 & 4756 , 1.4 & 4208 , 1.1 \\				\hline
			& &  \texttt{cumSeg}   			 & 5009 & 4928 , 1.6 & 4726 , 1.4 & 3893 , 0.9 \\				\hline
			& &  \texttt{FDRSeg} 			 & 5001 & 4953 , 1.6 & 4772 , 1.3 & 4253 , 1.1 \\				\hline			
			\hline
			& E.~(mix)  & \texttt{mscp}    &4936 &4929, 0.6& 4903, 0.5& 4707, 0.4\\		 \hline
			& &  \texttt{mosum} $G=50$   			 & 5040 & 4971 , 0.9 & 4848 , 0.7 & 4414 , 0.4 \\				\hline
			& &  \texttt{mosum} $G=100$	  		 & 4996 & 4969 , 0.8 & 4861 , 0.7 & 4472 , 0.4 \\				\hline
			& &  \texttt{mosum} $G=200$	   		 & 4001 & 3973 , 0.4 & 3912 , 0.3 & 3750 , 0.2 \\				\hline
			& &  \texttt{wbs}   			 & 6406 & 5056 , 0.8 & 4926 , 0.6 & 4547 , 0.4 \\				\hline
			& &  \texttt{not} '\texttt{MeanVar}'  			 & 5180 & 4995 , 0.8 & 4876 , 0.6 & 4504 , 0.4 \\			\hline
			& &  \texttt{not} '\texttt{MeanHT}'  			 & 5050 & 4987 , 0.8 & 4894 , 0.6 & 4519 , 0.4 \\			\hline
			& &  \texttt{changepoint}	   		 & 76448 & 11544 , 3.9 & 7751 , 1.9 & 5030 , 0.8 \\				\hline
			& &  \texttt{stepR}	  		 & 6288 & 5020 , 1.4 & 4905 , 1.2 & 4481 , 1 \\				\hline
			& &  \texttt{cumSeg}   			 & 5007 & 4938 , 1.5 & 4819 , 1.3 & 4034 , 0.9 \\				\hline
			& &  \texttt{FDRSeg} 			 & 5008 & 4984 , 1.3 & 4909 , 1.2 & 4499 , 1 \\				\hline		
			\hline
		\end{tabular}
	\end{scriptsize}
	\vspace{0.2cm}
	\caption{Comparative simulation study w.r.t.~scenario 1c.}
	\label{tab:comparison_1a}
\end{table}

\begin{table}[h!]
	\begin{scriptsize}
		\begin{tabular}{| l | l | l || l | l | l | l |}
			\hline
			Scn. & Distr. & Method & $|\hat C_{T}|$ & $|\hat C_{ 10}|$, $M_{10}$  & $|\hat C_{5}|$, $M_{5}$ & $|\hat C_{ 2}|$, $M_{2}$\\
			\hline	 \hline
			2c & A.~(normal) & \texttt{mscp}  &4884 &4873, 0.5 &4855, 0.5 &4663, 0.4\\	\hline
			& &  \texttt{mosum} $G=50$   			 & 5067 & 4964 , 0.9 & 4798 , 0.7 & 4348 , 0.4 \\				\hline
			& &  \texttt{mosum} $G=100$	  		 & 5015 & 4992 , 0.4 & 4935 , 0.3 & 4749 , 0.2 \\				\hline
			& &  \texttt{mosum} $G=200$	   		 & 3899 & 1505 , 2.3 & 1266 , 1.3 & 994 , 0.6 \\				\hline
			& &  \texttt{wbs}   			 & 5021 & 4985 , 0.7 & 4872 , 0.6 & 4539 , 0.4 \\				\hline
			& &  \texttt{not} '\texttt{MeanVar}'  			 & 5019 & 4991 , 0.7 & 4869 , 0.6 & 4532 , 0.4 \\				\hline
			& &  \texttt{not} '\texttt{MeanHT}'  			 & 5008 & 4977 , 0.8 & 4849 , 0.6 & 4484 , 0.4 \\				\hline
			& &  \texttt{changepoint}	   		 & 5022 & 4996 , 0.8 & 4887 , 0.6 & 4518 , 0.4 \\				\hline
			& &  \texttt{stepR}	  		 & 5046 & 4979 , 1.4 & 4859 , 1.2 & 4420 , 1 \\				\hline
			& &  \texttt{cumSeg}   			 & 5063 & 4890 , 1.7 & 4633 , 1.4 & 3750 , 0.9 \\				\hline
			& &  \texttt{FDRSeg} 			 & 5000 & 4979 , 1.4 & 4860 , 1.2 & 4422 , 1 \\				\hline
			\hline
			& B.~(gamma)& \texttt{mscp} &4906 &4870, 0.4 &4863, 0.3 &4766, 0.3\\ \hline
			& &  \texttt{mosum} $G=50$   			 & 5073 & 4881 , 1 & 4659 , 0.6 & 4232 , 0.3 \\				\hline
			& &  \texttt{mosum} $G=100$	  		 & 5048 & 4938 , 0.5 & 4850 , 0.3 & 4646 , 0.2 \\				\hline
			& &  \texttt{mosum} $G=200$	   		 & 3873 & 1475 , 1.9 & 1269 , 1 & 1060 , 0.4 \\				\hline
			& &  \texttt{wbs}   			 & 30508 & 6790 , 2.1 & 5658 , 0.9 & 4859 , 0.4 \\				\hline
			& &  \texttt{not} '\texttt{MeanVar}'  			 & 15290 & 4989 , 2 & 4217 , 0.9 & 3614 , 0.3 \\				\hline
			& &  \texttt{not} '\texttt{MeanHT}'  			 & 5041 & 4985 , 0.6 & 4917 , 0.5 & 4662 , 0.3 \\				\hline
			& &  \texttt{changepoint}	   		 & 14300 & 5495 , 2.5 & 4407 , 1.2 & 3420 , 0.4 \\				\hline
			& &  \texttt{stepR}	  		 & 32218 & 6701 , 2.7 & 5464 , 1.5 & 4518 , 1 \\				\hline
			& &  \texttt{cumSeg}   			 & 5063 & 4850 , 1.7 & 4603 , 1.4 & 3792 , 0.9 \\				\hline
			& &  \texttt{FDRSeg} 			 & 7008 & 4965 , 1.5 & 4763 , 1.3 & 4331 , 1 \\				\hline
			\hline
			& C.~(Poisson) & \texttt{mscp} & 4553 &4541, 0.7 &4520, 0.6 &4285, 0.5\\ \hline
			& &  \texttt{mosum} $G=50$   			 & 4820 & 4682 , 1.1 & 4482 , 0.8 & 3984 , 0.5 \\				\hline
			& &  \texttt{mosum} $G=100$	  		 & 4741 & 4681 , 0.5 & 4609 , 0.4 & 4396 , 0.2 \\				\hline
			& &  \texttt{mosum} $G=200$	   		 & 3394 & 1354 , 1.9 & 1193 , 1.2 & 976 , 0.6 \\				\hline
			& &  \texttt{wbs}   			 & 10456 & 5351 , 1.5 & 4907 , 0.9 & 4270 , 0.5 \\				\hline
			& &  \texttt{not} '\texttt{MeanVar}'  			 & 5215 & 4930 , 1.1 & 4729 , 0.8 & 4228 , 0.5 \\				\hline
			& &  \texttt{not} '\texttt{MeanHT}'  			 & 5094 & 4893 , 1.2 & 4662 , 0.8 & 4136 , 0.5 \\				\hline
			& &  \texttt{changepoint}	   		 & 5000 & 4966 , 0.9 & 4818 , 0.7 & 4361 , 0.4 \\				\hline
			& &  \texttt{stepR}	  		 & 9851 & 5223 , 2.1 & 4718 , 1.5 & 3916 , 1.1 \\				\hline
			& &  \texttt{cumSeg}   			 & 5039 & 4791 , 1.9 & 4487 , 1.6 & 3488 , 1 \\				\hline
			& &  \texttt{FDRSeg} 			 & 5742 & 4998 , 1.8 & 4689 , 1.4 & 3992 , 1.1 \\				\hline
			\hline
			& D.~(binomial) & \texttt{mscp} & 4847 &4841, 0.5 &4828, 0.5 &4647, 0.4\\ \hline
			& &  \texttt{mosum} $G=50$   			 & 4835 & 4728 , 0.9 & 4600 , 0.7 & 4173 , 0.4 \\				\hline
			& &  \texttt{mosum} $G=100$	  		 & 4799 & 4757 , 0.4 & 4711 , 0.3 & 4538 , 0.2 \\				\hline
			& &  \texttt{mosum} $G=200$	   		 & 3415 & 1389 , 1.8 & 1247 , 1.1 & 1049 , 0.6 \\				\hline
			& &  \texttt{wbs}   			 & 5984 & 5004 , 0.9 & 4837 , 0.7 & 4430 , 0.4 \\				\hline
			& &  \texttt{not} '\texttt{MeanVar}'  			 & 5168 & 4957 , 0.9 & 4807 , 0.7 & 4426 , 0.4 \\				\hline
			& &  \texttt{not} '\texttt{MeanHT}'  			 & 5076 & 4949 , 1 & 4775 , 0.7 & 4349 , 0.4 \\				\hline
			& &  \texttt{changepoint}	   		 & 195010 & 21834 , 4.9 & 12190 , 2.5 & 6114 , 1 \\				\hline
			& &  \texttt{stepR}	  		 & 5904 & 4988 , 1.6 & 4786 , 1.3 & 4219 , 1 \\				\hline
			& &  \texttt{cumSeg}   			 & 5030 & 4864 , 1.7 & 4615 , 1.4 & 3748 , 0.9 \\				\hline
			& &  \texttt{FDRSeg} 			 & 5034 & 4953 , 1.5 & 4803 , 1.3 & 4245 , 1 \\				\hline
			\hline
			& E.~(mix) & \texttt{mscp}&4926 &4920, 0.5 &4902, 0.5 &4720, 0.4\\ \hline
			& &  \texttt{mosum} $G=50$   			 & 5042 & 4963 , 0.8 & 4859 , 0.7 & 4469 , 0.4 \\				\hline
			& &  \texttt{mosum} $G=100$	  		 & 5026 & 4991 , 0.4 & 4950 , 0.3 & 4794 , 0.2 \\				\hline
			& &  \texttt{mosum} $G=200$	   		 & 3822 & 1551 , 2 & 1357 , 1.2 & 1110 , 0.6 \\				\hline
			& &  \texttt{wbs}   			 & 5880 & 5058 , 0.8 & 4954 , 0.6 & 4600 , 0.4 \\				\hline
			& &  \texttt{not} '\texttt{MeanVar}'  			 & 5113 & 4990 , 0.7 & 4898 , 0.6 & 4542 , 0.4 \\				\hline
			& &  \texttt{not} '\texttt{MeanHT}'  			 & 5022 & 4981 , 0.8 & 4883 , 0.7 & 4514 , 0.4 \\				\hline
			& &  \texttt{changepoint}	   		 & 40166 & 11549 , 3.9 & 7748 , 1.9 & 5000 , 0.8 \\				\hline
			& &  \texttt{stepR}	  		 & 5863 & 5042 , 1.4 & 4920 , 1.2 & 4484 , 1 \\				\hline
			& &  \texttt{cumSeg}   			 & 5053 & 4895 , 1.7 & 4682 , 1.4 & 3835 , 0.9 \\				\hline
			& &  \texttt{FDRSeg} 			 & 5044 & 4996 , 1.3 & 4917 , 1.2 & 4510 , 1 \\				\hline		
			\hline
		\end{tabular}
	\end{scriptsize}
	\vspace{0.2cm}
	\caption{Comparative simulation study w.r.t.~scenario 2c.}
	\label{tab:comparison_2a}
\end{table}

\begin{table}[h!]
	\begin{scriptsize}
		\begin{tabular}{| l | l | l || l | l | l | l |}
			\hline
			Scn. & Distr. & Method & $|\hat C_{T}|$ & $|\hat C_{ 10}|$, $M_{10}$  & $|\hat C_{5}|$, $M_{5}$ & $|\hat C_{ 2}|$, $M_{2}$\\
			\hline	 \hline
			3c & A.~(normal)  & \texttt{mscp} &4814& 4703, 1.3& 4286, 0.7& 3936, 0.4\\ \hline
			& &  \texttt{mosum} $G=50$   			 & 5063 & 4955 , 0.8 & 4823 , 0.6 & 4452 , 0.3 \\				\hline
			& &  \texttt{mosum} $G=100$	  		 & 4344 & 2665 , 1.7 & 2445 , 1.2 & 2008 , 0.6 \\				\hline
			& &  \texttt{mosum} $G=200$	   		 & 3055 & 2032 , 0.8 & 1983 , 0.6 & 1823 , 0.4 \\				\hline
			& &  \texttt{wbs}   			 & 5018 & 4975 , 0.7 & 4873 , 0.6 & 4519 , 0.4 \\				\hline
			& &  \texttt{not} '\texttt{MeanVar}'  			 & 5009 & 4980 , 0.8 & 4866 , 0.6 & 4510 , 0.4 \\				\hline
			& &  \texttt{not} '\texttt{MeanHT}'  			 & 5012 & 4973 , 0.8 & 4832 , 0.6 & 4423 , 0.4 \\				\hline
			& &  \texttt{changepoint}	   		 & 5026 & 5003 , 0.8 & 4894 , 0.6 & 4532 , 0.4 \\				\hline
			& &  \texttt{stepR}	  		 & 5045 & 4973 , 1.4 & 4855 , 1.2 & 4445 , 1 \\				\hline
			& &  \texttt{cumSeg}   			 & 5086 & 4831 , 1.7 & 4587 , 1.4 & 3824 , 0.9 \\				\hline
			& &  \texttt{FDRSeg} 			 & 5000 & 4976 , 1.4 & 4860 , 1.2 & 4449 , 1 \\				\hline
			\hline
			& B.~(gamma) & \texttt{mscp}   &4820& 4749, 1.2& 4334, 0.5& 4095, 0.3\\	\hline
			& &  \texttt{mosum} $G=50$   			 & 5096 & 4903 , 0.8 & 4719 , 0.6 & 4360 , 0.3 \\				\hline
			& &  \texttt{mosum} $G=100$	  		 & 4377 & 2572 , 1.7 & 2310 , 1 & 1938 , 0.5 \\				\hline
			& &  \texttt{mosum} $G=200$	   		 & 3076 & 2023 , 0.9 & 1942 , 0.6 & 1780 , 0.3 \\				\hline
			& &  \texttt{wbs}   			 & 23085 & 6251 , 1.7 & 5425 , 0.8 & 4767 , 0.4 \\				\hline
			& &  \texttt{not} '\texttt{MeanVar}'  			 & 13558 & 5090 , 1.9 & 4344 , 0.9 & 3744 , 0.4 \\				\hline
			& &  \texttt{not} '\texttt{MeanHT}'  			 & 5021 & 4988 , 0.6 & 4927 , 0.5 & 4667 , 0.3 \\				\hline
			& &  \texttt{changepoint}	   		 & 12527 & 5541 , 2.6 & 4437 , 1.3 & 3411 , 0.4 \\				\hline
			& &  \texttt{stepR}	  		 & 23771 & 6141 , 2.4 & 5188 , 1.4 & 4451 , 1 \\				\hline
			& &  \texttt{cumSeg}   			 & 5085 & 4812 , 1.6 & 4598 , 1.3 & 3859 , 0.8 \\				\hline
			& &  \texttt{FDRSeg} 			 & 7424 & 5046 , 1.7 & 4746 , 1.3 & 4327 , 1 \\				\hline
			\hline
			& C.~(Poisson) & \texttt{mscp}    &4387& 4249, 1.5& 3845, 0.8& 3480, 0.5\\	\hline
			& &  \texttt{mosum} $G=50$   			 & 4860 & 4703 , 0.9 & 4531 , 0.6 & 4133 , 0.3 \\				\hline
			& &  \texttt{mosum} $G=100$	  		 & 3605 & 2295 , 1.8 & 2075 , 1.2 & 1693 , 0.6 \\				\hline
			& &  \texttt{mosum} $G=200$	   		 & 3021 & 1989 , 1 & 1904 , 0.8 & 1731 , 0.5 \\				\hline
			& &  \texttt{wbs}   			 & 9732 & 5336 , 1.5 & 4889 , 0.9 & 4223 , 0.5 \\				\hline
			& &  \texttt{not} '\texttt{MeanVar}'  			 & 5226 & 4910 , 1.1 & 4683 , 0.8 & 4184 , 0.5 \\				\hline
			& &  \texttt{not} '\texttt{MeanHT}'  			 & 5039 & 4901 , 1.2 & 4670 , 0.9 & 4122 , 0.5 \\				\hline
			& &  \texttt{changepoint}	   		 & 5002 & 4975 , 0.9 & 4846 , 0.7 & 4408 , 0.4 \\				\hline
			& &  \texttt{stepR}	  		 & 9140 & 5176 , 2.1 & 4660 , 1.5 & 3894 , 1.1 \\				\hline
			& &  \texttt{cumSeg}   			 & 5027 & 4559 , 1.7 & 4346 , 1.5 & 3561 , 1 \\				\hline
			& &  \texttt{FDRSeg} 			 & 5928 & 5156 , 2 & 4724 , 1.5 & 3986 , 1.1 \\				\hline
			\hline
			& D.~(binomial)& \texttt{mscp} &4750& 4644, 1.5& 4146, 0.7& 3809, 0.4\\	\hline
			& &  \texttt{mosum} $G=50$   			 & 4870 & 4786 , 0.8 & 4655 , 0.6 & 4284 , 0.3 \\				\hline
			& &  \texttt{mosum} $G=100$	  		 & 3703 & 2347 , 1.6 & 2160 , 1.1 & 1819 , 0.6 \\				\hline
			& &  \texttt{mosum} $G=200$	   		 & 2974 & 1970 , 0.8 & 1935 , 0.6 & 1796 , 0.4 \\				\hline
			& &  \texttt{wbs}   			 & 5920 & 5006 , 0.9 & 4837 , 0.7 & 4409 , 0.4 \\				\hline
			& &  \texttt{not} '\texttt{MeanVar}'  			 & 5113 & 4962 , 0.9 & 4819 , 0.7 & 4405 , 0.4 \\				\hline
			& &  \texttt{not} '\texttt{MeanHT}'  			 & 5052 & 4955 , 1 & 4778 , 0.7 & 4337 , 0.4 \\				\hline
			& &  \texttt{changepoint}	   		 & 185641 & 21685 , 4.9 & 12126 , 2.5 & 6151 , 1 \\				\hline
			& &  \texttt{stepR}	  		 & 5851 & 4962 , 1.6 & 4761 , 1.4 & 4206 , 1 \\				\hline
			& &  \texttt{cumSeg}   			 & 5093 & 4690 , 1.6 & 4515 , 1.4 & 3756 , 0.9 \\				\hline
			& &  \texttt{FDRSeg} 			 & 5093 & 4967 , 1.5 & 4792 , 1.3 & 4262 , 1 \\				\hline
			\hline
			& E.~(mix) & \texttt{mscp}     &4856& 4756, 1.3& 4334, 0.7& 3990, 0.4\\ 	\hline			
			& &  \texttt{mosum} $G=50$   			 & 5060 & 4983 , 0.7 & 4866 , 0.6 & 4538 , 0.3 \\				\hline
			& &  \texttt{mosum} $G=100$	  		 & 4339 & 2650 , 1.6 & 2446 , 1.1 & 2052 , 0.6 \\				\hline
			& &  \texttt{mosum} $G=200$	   		 & 3059 & 2045 , 0.8 & 1992 , 0.7 & 1841 , 0.4 \\				\hline
			& &  \texttt{wbs}   			 & 6026 & 5082 , 0.8 & 4957 , 0.7 & 4581 , 0.4 \\				\hline
			& &  \texttt{not} '\texttt{MeanVar}'  			 & 5094 & 4991 , 0.8 & 4895 , 0.6 & 4547 , 0.4 \\				\hline
			& &  \texttt{not} '\texttt{MeanHT}'  			 & 5012 & 4982 , 0.8 & 4883 , 0.6 & 4524 , 0.4 \\				\hline
			& &  \texttt{changepoint}	   		 & 21991 & 11491 , 3.8 & 7769 , 1.9 & 5059 , 0.8 \\				\hline
			& &  \texttt{stepR}	  		 & 5938 & 5052 , 1.4 & 4914 , 1.3 & 4461 , 1 \\				\hline
			& &  \texttt{cumSeg}   			 & 5069 & 4818 , 1.6 & 4620 , 1.4 & 3833 , 0.9 \\				\hline
			& &  \texttt{FDRSeg} 			 & 5174 & 5030 , 1.4 & 4915 , 1.2 & 4485 , 1 \\				\hline		
			\hline
		\end{tabular}
	\end{scriptsize}
	\vspace{0.2cm}
	\caption{Comparative simulation study w.r.t.~scenario 3c.} 
	\label{tab:comparison_3a}
\end{table}

\begin{figure}[h!]
	\centering  
	\includegraphics[width=0.7\textwidth,angle=0]{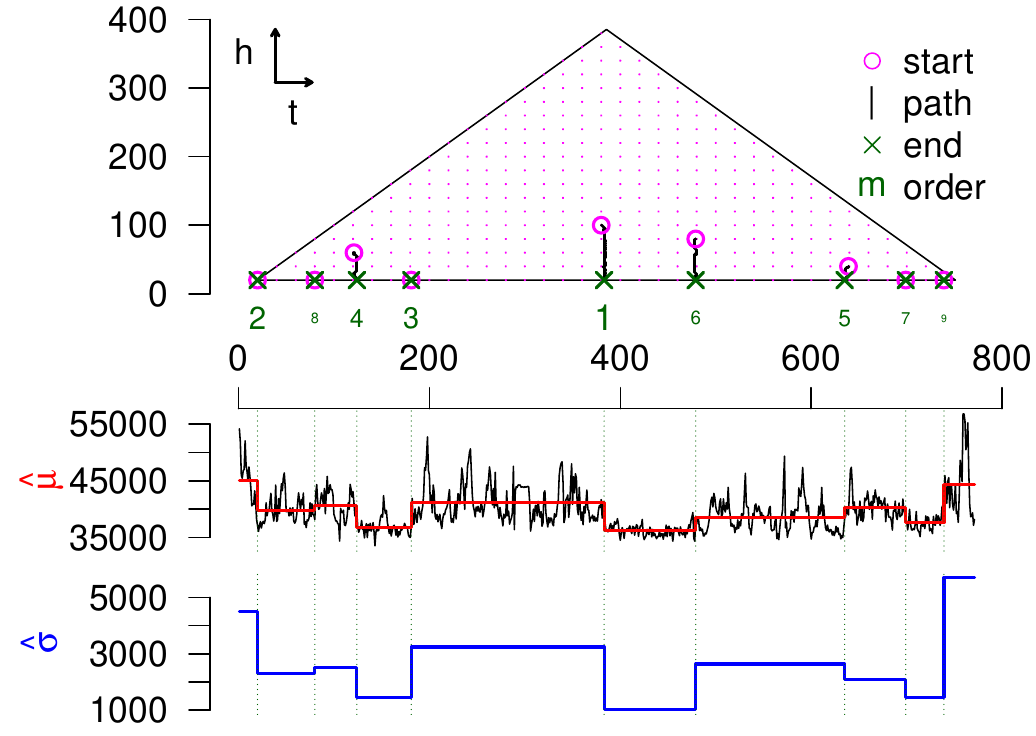}
	\caption{Analysis of the base cytosine in the X-chromosome of T2T-CHM13.}
	\label{fig:human}
\end{figure}

\paragraph{Data example}
The complete sequence of a human genome (T2T-CHM13) was recently presented for the first time in \citet{Nurk2022}. We analyze the nucleotide sequence of the X-chromosome (Homo sapiens isolate CHM13 chromosome X, GenBank: CP068255.2). It consists of about $b=1542\cdot10^5$ base pairs. We segment the sequence into segments of length $d=2\cdot10^5$, within which we compute the frequency of the base cytosine. This yields $T=b/d=771$ data points to which we apply MSCP, see Figure \ref{fig:human}.

Nine change points were estimated, $\hat C=\{20 , 80 ,124, 181, 383, 479, 635, 699, 739\}$, yielding paramter estimates 
$\hat\mu_u\approx 
(45, 40, 41, 37, 41, 36, 39, 40, 38, 44) \cdot 10^3$
and \newline
$\hat \sigma_u\approx 
(4.5 , 2.3, 2.5, 1.5, 3.2, 1.0, 2.6, 2.1, 1.5, 5.7)\cdot 10^3$.
The segmentation closely aligns with visual inspection. We note that model assumptions are fairly met: first, an increase in mean is accompanied with an increase in variance, and vice versa, which aligns with the assumption that the theoretical variance may also change when change in expectation occurs. Second, in each section, serial correlation proves to be moderate (e.g., for lag one it is between $0.3$ (section 6) and $0.7$ (last section 10), and it rapidly decreases for higher lags). Third, the representation of the data as frequency counts is supported by the non-parametric assumptions on the distributions. Finally, we mention that the result is stable under variation of $d$. Being an interesting result in itself, the example shows that MSCP can be helpful for the segmentation of genomic data: yielding segments with relatively stable first order moments, it can be considered a preprocessing step in data analysis.  

	\section{Appendix}\label{sect:appendix}
	\noindent\textbf{Proof of Lemma \ref{conv:est_cp}}:  
	W.l.o.g.~let $j=\ell$. For $\hat\mu_\ell^{\langle k \rangle}$ first let $C=\emptyset$ where we need to show $n^v\cdot (\hat{\mu}_\ell^{\langle k \rangle} -\mu^{\langle k\rangle})_{(t,h)}\to (0)_{(t,h)}$ a.s.~as $n\to\infty$. We show some uniform convergence w.r.t~$[0,T]$ and then extend considerations to $\Delta_\delta$. It holds a.s.~as $n\to\infty$ 
	\begin{equation}\label{fslln}
	\begin{split}
	n^v\cdot &\sup_{t\in[0,T]}\left| 
	\frac{1}{n}\sum_{i=1}^{\lfloor nt\rfloor} X_i^k -t\mu^{\langle k \rangle} \right| \longrightarrow 0, \\
	&\qquad \qquad\qquad\qquad\mathrm{for}\; v\in
	\begin{cases}
	(-\infty,1/2),&\textrm{if}\; k=1,\\
	(-\infty,1/2)\cap(-\infty,p/(p+2)],&\textrm{if}\; k=2.
	\end{cases}	
	\end{split}
	\end{equation}
	
	For this, first apply Marcinkiewicz-Zygmund (MZ)-SLLN to $(X_i)_i$ and $(X_i^2)_i$: For $(X_i)_i$ note that for any $\alpha\in(0,2)$ it is $\ew[|X_i|^\alpha]<\infty$, and thus for $k=1$
	a.s.~as $n\to\infty$
	\begin{align}\label{MZSLLN1}
	n^{(\alpha-1)/\alpha} \cdot \Big( \frac{1}{n} \sum_{i=1}^n X_i^k-\mu^{\langle k\rangle}\Big)
	=
	n^{-1/\alpha} \cdot \Big(\sum_{i=1}^n X_i^k-n\mu^{\langle k\rangle}\Big)\longrightarrow 0.
	\end{align} 
	For the squares $(X_i^2)_i$ we choose $\alpha\in (0,(2+p)/2]$ if $p<2$ and $\alpha\in (0,2)$ if $p\ge2$. Then $\ew[|X_i^2|^\alpha]<\infty$ and thus (\ref{MZSLLN1}) applies for $k=2$. As $(\alpha-1)/\alpha\le p/(2+p)$ if $p<2$, and $(\alpha-1)/\alpha < 1/2$ if $p\ge 2$, we obtain for $k=1,2$ a.s.~as $n\to\infty$
	\begin{align}\label{rate_null1X}
	n^v\cdot \Big( \frac{1}{n} \sum_{i=1}^n X_i^k-\mu^{\langle k\rangle}\Big)&\longrightarrow 0,
	\;\, \mathrm{for}\; v\in
	\begin{cases}
	(-\infty,1/2),& \textrm{if}\; k=1,\\
	(-\infty,1/2)\cap(-\infty,p/(p+2)],&\textrm{if}\; k=2.
	\end{cases}	
	\end{align}	
	
	For uniform convergence in (\ref{fslln}) we discretize time and apply (\ref{rate_null1X}), a.s.~as $n\to\infty$
	\begin{align*}
	n^v\cdot\sup_{0\le t\le T}\left|\frac{1}{n}\sum_{i=1}^{\lfloor nt \rfloor} X_i^k - t\mu^{\langle k\rangle}\right|
	&\!\le\!
	n^v\cdot \sup_{0\le t\le T}\left|\frac{1}{n}\sum_{i=1}^{\lfloor nt \rfloor} X_i^k  - \frac{\lfloor nt\rfloor}{n}\mu^{\langle k\rangle}\right| + r_n\!\\
	&= n^v\cdot \!\!\!\max_{m \in\{1, 2,\ldots,nT\}}\left|\frac{1}{n}\sum_{i=1}^{m} X_i^k - \frac{m}{n}\mu^{\langle m\rangle} \right|\! + r_n \longrightarrow 0, 
	\end{align*}
	as $r_n := n^v\cdot \sup_{t\in[0,T]}|t - \lfloor nt \rfloor/n| \; \mu^{\langle k\rangle}\le n^v\cdot(1/n) \;\mu^{\langle k\rangle}\to 0$ as $n\to\infty$, and for $\varepsilon >0$
	\begin{align*}
	&n^v\cdot\max_{m \in\{1, 2,\ldots,nT\}}\left|\frac{1}{n}\sum_{i=1}^{m} X_i^k - \frac{m}{n}\mu^{\langle k\rangle} \right| \le \\
	& n^v\cdot \frac{1}{n} \max_{m\le m_0}\left|\sum_{i=1}^m X_i^k -m\mu^{\langle k\rangle}\right| 
	+ T^{1-v}\max_{m_0<m\le nT} \Big(\frac{m}{nT}\Big)^{1-v} m^v\cdot\left|\frac{1}{m}\sum_{i=1}^m X_i^k -\mu^{\langle k\rangle}\right|\le \varepsilon 
	\end{align*}
	a.s.~for $n$ large enough, as the maximum in the first summand is independent from $n$, and for the second summand note $(m/nT)^{1-v}\le1$ and  
	(\ref{rate_null1X}) yields that for almost every realization we find $m_0\in\nn$ such that $m^v\cdot|(1/m)\sum_{i=1}^m X_i^k - \mu^{\langle k\rangle}|<T^{v-1}\varepsilon/2$ for all $m> m_0$. Thus, (\ref{fslln}) holds. Now we consider $\Delta_\delta$.
	As $0\le t-h \le T$ for all $(t,h)\in\Delta_\delta$, (\ref{fslln}) yields a.s.~as $n\to\infty$
	\begin{align*}
	&n^v\cdot \sup_{(t,h)\in\Delta_\delta} \left|\frac{1}{n}\sum_{i=1}^{\lfloor nt \rfloor} X_i^k - t\mu^{\langle k\rangle}\right| \longrightarrow 0
	\quad \textrm{and} \quad\\
	&n^v\cdot \sup_{(t,h)\in\Delta_\delta} \left|\frac{1}{n}\sum_{i=1}^{\lfloor n(t-h) \rfloor} X_i^k - (t-h)\mu^{\langle k\rangle}\right| 
	\longrightarrow 0,
	\end{align*}
	We include the factor $1/h\ge2/T>0$ and obtain in  $(\mathcal D_\rn [\Delta_\delta],\|\cdot\|_\infty)$ a.s.~as $n\to\infty$
	\begin{align}\label{conv:aux2}
	&n^v \cdot (\hat{\mu}_\ell^{\langle k \rangle} - \mu^{\langle k\rangle})_{(t,h)}=\\
	&n^v \cdot
	\Big[
	\Big(
	\frac{1}{nh} \sum\nolimits_{i=1}^{\lfloor nt\rfloor} X_i^k - \frac{t }{h}\mu^{\langle k\rangle}
	\Big)
	- 
	\Big(
	\frac{1}{nh} \sum\nolimits_{i=1}^{\lfloor n(t-h)\rfloor} X_i^k - \frac{t-h }{h}\mu^{\langle k\rangle}
	\Big)
	\Big]_{(t,h)}
	\rightarrow (0)_{(t,h)}.\nonumber
	\end{align}
	Now let $C\not=\emptyset$. We segment $(t-h,t]$ according to $C$ and find $\hat{\mu}_\ell^{\langle k \rangle} - \mu^{\langle k\rangle}$ as
	\begin{align}\label{conv:mupoint}
	\sum_{u=1}^{|C_{\ell}|+1}
	\Big(
	\Big[
	\frac{\lfloor nc_{\ell,u} \rfloor-\lfloor nc_{\ell,u-1} \rfloor}{nh} \cdot 
	\frac{1}{ \lfloor nc_{\ell,u} \rfloor-\lfloor nc_{\ell,u-1} \rfloor} \sum_{i=\lfloor nc_{\ell,u-1} \rfloor+1}^{ \lfloor nc_{\ell,u} \rfloor} X_i^k
	\Big]
	-
	\frac{d_{\ell,u}}{h} \cdot \mu_{\ell,u}^{\langle k\rangle}
	\Big).
	\end{align}
	The $u$-th summand refers to a subsection that relates to the error sequence $(Z_{u,i})_{i=1,2,\ldots}$ and in which $X_i$ equals $\mu_u +\sigma_u \cdot Z_{u,i}$. From (\ref{conv:aux2}) we conclude a.s.~as $n\to\infty$
	\begin{align}\label{varyingd}
	n^v\cdot	\sup_{1\le d\le t \le T}\left| \frac{1}{\lfloor nd \rfloor } \sum_{i=\lfloor n(t-d)\rfloor +1}^{\lfloor nt \rfloor} (\mu_u +\sigma_u \cdot Z_{u,i})^k - \mu_u^{\langle k \rangle} \right| \longrightarrow 0,
	\end{align}
	which states uniform convergence w.r.t~all subintervals of varying length $d$. When including the factor $d/h\ge 1$ and summing over all $u$, the expression still vanishes a.s.~as $n\to\infty$, and it states an upper bound for $n^v\cdot\sup_{(t,h)\in\Delta_\delta}|\hat\mu_\ell^{\langle k \rangle} - \tilde \mu_\ell^{\langle k \rangle}|$.
	
	Regarding $\hat\sigma_\ell^2$ we decompose in the $u$-th subsection
	$(X_{i}-\hat\mu_{\ell})^2 = (X_{i}-\mu_{\ell,u})^2-2(X_{i}-\mu_{\ell,u})(\hat\mu_\ell-\mu_{\ell,u}) + (\hat\mu_\ell-\mu_{\ell,u})^2$. Averages in the subsection tend to $\sigma_{\ell,u}^2 + 0 +  (\tilde\mu_{\ell}-\mu_{\ell,u})^2$, and summation over subsections yields a.s.~as $n\to\infty$ that $n^v\cdot\sup_{(t,h)\in\Delta_\delta}|\hat\sigma_\ell^{2} - \tilde \sigma_\ell^{2}|\to 0$, as before using MZ-SLLN and discretization arguments.\hfill\qed\\

	\noindent
	\textbf{Proof of Lemma \ref{lemm_shark}}: 
	Continuity is inherited from the limits in (\ref{tilde_left}) and (\ref{error}). W.l.o.g.~consider $c_1=:c$. As $v_{t,h}\ge 1$ with equality at $c$ it is $v_{t,h}\cdot| d_{t,h}^{(n)}|\ge |d_{t,h}^{(n)}|$.
	$\tilde\mu_r-\tilde\mu_\ell$ has the shape of a hat: it is zero outside the $h$-neighborhood of $c$, it is $\mu_{2}-\mu_1$ at $c$, and it is linearly interpolated in between. $\tilde\tilde\sigma_r^2+\tilde\tilde\sigma_\ell^2$ is $2\sigma_1^2$ left of $c-h$, and $2\sigma_{2}^2$ right of $c+h$, and linearly interpolated in between. Thus, in the $h$-neighborhood of $c$ the root $(\tilde\tilde\sigma_r^2+\tilde\tilde\sigma_\ell^2)^{1/2}$ is constant if $\sigma_{2}^2 = \sigma_1^2$, it
	is strictly convex if $\sigma_{2}^2 > \sigma_1^2$ and strictly concave if $\sigma_{2}^2 < \sigma_1^2$. As the numerator is piecewise linear, i.e., of order $t$, and the denominator of order $t^{1/2}$, the statements about the curvatures of $v_{t,h}\cdot d_{t,h}^{(n)}$ hold true.
	We now turn to $d_{t,h}^{(n)}$. We represent $\tilde\sigma_j^2$ through $\tilde\tilde\sigma_j^2$ plus errors 
	\[
	\tilde\sigma_{j}^2 = 
	\sum_{u=1}^{2} \frac{d_u}{h} \cdot [\sigma_u^2 +(\tilde \mu_j -  \mu_u)^2]
	= \tilde\tilde\sigma_j^2 + \frac{d_1d_2}{h^2}\cdot(\mu_2-\mu_1)^2,
	\]
	for $j\in \{\ell,r\}$, i.e., we find the error as $e_j^2:=[d_1d_2/h^2](\mu_2-\mu_1)^2$. We abbreviated $d_u:=d_{j,u}$.  Note that $e_\ell^2=e_r^2$,
	and set $x:=d_1/h$ and $d_2/h=(1-h)/h=(1-x)$ with $x\in[0,1]$, which yields a representation through the proportions of the window left and right of $c$. Note that the error $e_j^2=x(1-x)\cdot(\mu_2-\mu_1)^2$ is quadratic and maximal for $x=1/2$, thus taking the value $(\mu_2-\mu_1)^2/4$, which is plausible as half of the window refers to the left and the other half to the right population. Within both $[c-h,c]$ and $(c,c+h]$ it holds that $\tilde\mu_r-\tilde\mu_\ell$ is linear in $x$ and the denominator of $d_{t,h}^{(n)}$ is now the root of first function $\tilde\tilde\sigma_r^2+\tilde\tilde\sigma_\ell^2$ which is linear in $x$ plus second the quadratic error $e_j^2$.
	W.l.o.g.~we consider $t\in[c-h,c]$ where the right window contains $c$.
	Then we find $d_{t,h}^{(n)}$ as a function $f$ of $x$ as
	\[
	f(x)
	= \sqrt{nh}\cdot \frac{(\mu_2-\mu_1) \cdot x}{\sqrt{[(\sigma_2^2-\sigma_1^2)\cdot x +2\sigma_1^2] + [(\mu_2-\mu_1)^2\cdot x\cdot(1-x)]}},\qquad x\in[0,1],
	\]
	for all valid $h\in (\delta,T/2]$, which yields the derivative w.r.t.~$x$
	\[
	f'(x) = \sqrt{nh}\cdot (\mu_2-\mu_1) \cdot \frac{2^{-1}[(\sigma_2^2-\sigma_1^2)+(\mu_2-\mu_1)^2]\cdot x + 2\sigma_1^2}{[(\sigma_2^2-\sigma_1^1)\cdot x+ 2\sigma_1^2 + (\mu_2-\mu_1)^2\cdot x\cdot (1-x)]^{3/2}}.
	\]
	In (\ref{lowerbound}) we see that the fraction is positive and thus $f'(x)>0$ if $\mu_2>\mu_1$ and $f'(x)<0$ if $\mu_2<\mu_1$, which gives (\ref{dshape}). To bound $|f'(x)|$ we find the numerator $\ge 2^{-1}[\sigma_2^2+ 3\sigma_1^2 + (\mu_2-\mu_1)^2]\wedge 2\sigma_1^2$ and for the denominator we get $(\sigma_2^2-\sigma_1^2)x+2\sigma_1^2\le 2(\sigma_2^2\vee \sigma_1^2)$ and $x(1-x)\le 1/4$, such that
	\begin{align}\label{lowerbound}
	|f'(x)| 
	&\ge \sqrt{nh}\cdot |\mu_2-\mu_1|\cdot \frac{2^{-1}[\sigma_2^2+ 3\sigma_1^2+(\mu_2-\mu_1)^2]\wedge 2\sigma_1^2}{[2(\sigma_2^2\vee \sigma_1^2) + (\mu_2-\mu_1)^2/4]^{3/2}}\nonumber\\
	&\ge 
	\sqrt{nh}\cdot |\mu_2-\mu_1|\cdot 
	\frac{2(\sigma_2^2\wedge \sigma_1^2)}{[2(\sigma_2^2\vee \sigma_1^2) + (\mu_2-\mu_1)^2/4]^{3/2}}>0.
	\end{align}
	In the second inequality we omitted $(\mu_2-\mu_1)^2 \ge 0$ and used that $2\sigma_{1}^2 <2^{-1}[\sigma_{2}^2+3\sigma_{1}^2]$ iff $\sigma_{1}^2<\sigma_2^2$. This is the lower bound in (\ref{derivative}).
	For the upper bound we find the numerator $\le 2^{-1}[\sigma_2^2+ 3\sigma_1^2 + (\mu_2-\mu_1)^2]\vee 2\sigma_1^2\le  
	2(\sigma_2^2\vee \sigma_1^2) +(\mu_2-\mu_1)^2$, and for the denominator we mention  $(\sigma_2^2-\sigma_1^2)x+2\sigma_1^2\ge 2(\sigma_2^2\wedge \sigma_1^2)$ such that
	\begin{align*}
	|f'(x)| 
	&\le
	\sqrt{nh}\cdot |\mu_2-\mu_1|\cdot 
	\frac{2(\sigma_2^2\vee \sigma_1^2) + (\mu_2-\mu_1)^2}{[2(\sigma_2^2\wedge \sigma_1^2)]^{3/2}},
	\end{align*}
	which completes (\ref{derivative}).
	\hfill\qed\\
	
	\noindent\textbf{Proof of Proposition \ref{prop_main}}: 
	Donsker's theorem yields in $(\mathcal D_{\rn}[0,T],\|\cdot\|_\infty)$ as $n\to\infty$ 
	\begin{align}\label{donsker}
	\Big[
	\frac{1}{\sigma\sqrt{n}}\cdot
	\sum\nolimits_{i=1}^{\lfloor nt \rfloor} (X_i-\mu)
	\Big]_{t}
	\stackrel{d}{\longrightarrow}
	(W_t)_{t}.
	\end{align}
	Define a continuous map $\varphi$ from $(\mathcal D_{\rn}[0,T],\|\cdot\|_\infty)$ to $(\mathcal D_{\rn}[\Delta_\delta],\|\cdot\|_\infty)$ via
	\[
	\varphi : (f(t))_{t} \to \left(\frac{[f(t+h)-f(t)]-[f(t)-f(t-h)]}{\sqrt{2h}}\right)_t,
	\]
	and apply $\varphi$ on (\ref{donsker}). The
	continuous mapping yields in $(\mathcal D_{\rn}[\Delta_\delta],\|\cdot\|_\infty)$ as $n\to\infty$ 
	\begin{align*}
	\left(
	\frac{1}{(2\sigma^2nh)^{1/2}}
	\left[\sum\nolimits_{i = \lfloor nt \rfloor +1 }^{ \lfloor n(t+h) \rfloor} X_i - \sum\nolimits_{i = \lfloor n(t-h) \rfloor +1}^{ \lfloor nt \rfloor} X_i\right]
	\right)_{t}
	\stackrel{d}{\longrightarrow}
	(L_{t,h})_t,
	\end{align*}
	where the constant $\mu$ vanishes. Now by replacing $2\sigma^2$ by $\hat\sigma_r^2 + \hat\sigma_\ell^2$ and using Lemma \ref{conv:est_cp} and Slutsky's theorem weak convergence of $(D_{t,h}^{(n)})_{(t,h)\in\Delta_\delta}$ follows. \hfill$\Box$\\ 	 
	
	\noindent\textbf{Proof of Lemma \ref{lemm:pathdth}}: Within $A_u$ the slope of $(n^{-1/2}\cdot|d_{t,h}^{(n)}|)_t$ is positive left of $c_u$ and negative right of it, see Lemma \ref{lemm_shark}. Thus, $|t_s(k)-c_u|=|t_s-c_u|-(k+1)\Leftrightarrow$ $|t_s-c_u|\ge k+1$ and (\ref{optimal:path}) holds true. The path ends after $h_s- \delta $ steps, i.e., $t_e=c_u\Leftrightarrow$ $|t_s-c_u|\le  h_s- \delta  + 1\Leftrightarrow$ $(t_s,h_s)\in B_u$. Further, $(t_s,h_s)\in A_u$ implies $|t_s-c_u|\le h_s$ (double window overlaps $c_u$), and thus $(t_s,h_s)\in A_u\backslash B_u$ implies $|t_s-c_u|\in \{h_s-\delta+2,\ldots,h_s\}$, hence $|t_e-c_u|=|t_s-c_u|-(h_s- \delta +1)\in\{1,\ldots, \delta\}$.\hfill $\Box$\\	
	
	\noindent{\textbf{Proof of Proposition \ref{lemm:pathDth}}}
	From Corollary \ref{conv:Dth} we obtain a.s.~as $n\to\infty$
	\begin{align}\label{cond:c1}
	&  \sup\nolimits_{(t,h)\in \Delta_\delta}\, \left| |n^{-1/2}\cdot D_{t,h}^{(n)}| -  |d_{t,h}^{(1)}| \right|  \longrightarrow 0.
	\end{align}
	On $A_u$ (excluding $t=c_u$) a lower bound for the derivative of $(|d_{t,h}^{(1)}|)_t$ is given through $\kappa_a\delta^{1/2}>0$, see Lemma \ref{lemm_shark}.
	Thus, for any $\varepsilon>0$ it is
	\begin{align}
	|d_{t,h}^{(1)}|-|d_{t-\varepsilon,h}^{(1)}|\ge\kappa_a\delta^{1/2}\varepsilon 
	,\; \textrm{if $t\le c_u$}, 
	\quad
	\textrm{and}		
	\quad 
	|d_{t,h}^{(1)}|-|d_{t+\varepsilon,h}^{(1)}|\ge \kappa_a\delta^{1/2}\varepsilon, \; \textrm{if $t> c_u$},\label{cond:c2}
	\end{align}
	provided that $(t,h)$, $(t-\varepsilon,h)$, $(t+\varepsilon,h)$ lie in $A_u$.
	Consider the initializing step of the path, $k=0$. Both $\hat t_s^{(n)}(k)$ and $t_s(k)$ take a value in $\{t_s-1,t_s,t_s+1\}\cap\Delta_\delta$. As $\hat t_s^{(n)}(k)$ is a maximizer defined via $|D_{t,h}^{(n)}|$ we obtain
	$
	|n^{-1/2} \cdot D_{\hat t_s^{(n)}(k),h-k}^{(n)}| \ge
	|n^{-1/2} \cdot D_{t_s(k),h-k}^{(n)}| \to
	| d_{t_s(k),h-k}^{(1)}|
	$ a.s.~as $n\to \infty$,
	where the convergence follows from \eqref{cond:c1},
	i.e.,
	\begin{align}\label{D_at_est}
	|n^{-1/2}\cdot D_{\hat t_s^{(n)}(k),h-k}^{(n)}| \ge  | d_{t_s(k),h-k}^{(1)}|  + o_{a.s.}(1).
	\end{align}
	From this we bound
	\begin{align}\label{est_to_d}
	| d_{t_s(k),h-k}^{(1)}| -  | d_{\hat t_s^{(n)}(k),h-k}^{(1)}| &+ o_{a.s.}(1) \nonumber\\
	&\le |n^{-1/2}\cdot D_{\hat t_s^{(n)}(k),h-k}^{(n)}| -  | d_{\hat t_s^{(n)}(k),h-k}^{(1)}| + o_{a.s}(1) \nonumber \\
	&\le \left| n^{-1/2}\cdot D_{\hat t_s^{(n)}(k),h-k}^{(n)} - d_{\hat t_s^{(n)}(k),h-k}^{(1)}  \right| + o_{a.s.}(1)\nonumber \\ 
	& \le \sup_{(t,h)\in\Delta_\delta} \left|  n^{-1/2}\cdot D_{t,h}^{(n)} - d_{t,h}^{(1)}  \right| + o_{a.s.}(1) 
	\stackrel{a.s.}{\longrightarrow} 0.
	\end{align}	
	In the first inequality we used \eqref{D_at_est}, in the second the triangle inequality, 
	and the convergence follows from \eqref{cond:c1}. The estimator $\hat t_s^{(n)}(k)$ is now tied to the nonrandom function via $d_{\hat t_s^{(n)}(k),h-k}^{(1)}$, and \eqref{cond:c2} brings us from the function to the estimator. Let $\varepsilon >0$. Then almost everywhere it holds that if $|t_s(k)-\hat t_s^{(n)}(k)|>\varepsilon$ i.o., thus $|d_{t_s(k),h-k}^{(1)}| - | d_{\hat t_s^{(n)}(k),h-k}^{(1)}| >\varepsilon\kappa_a \delta^{1/2}$ i.o. But as the latter only occurs finitely often it is $t_s(k) =\hat t_s^{(n)}(k)$ a.s.~for $n$ large. Iteratively, this extends for all $k=1,2,..,h- \delta$, i.e., eventually the paths w.r.t.~$D_{t,h}^{(n)}$ and $d_{t,h}^{(n)}$ coincide a.s.\hfill$\Box$\\
	
	\noindent\textbf{Proof of Lemma \ref{lemm:scapa}:}
	From $ \delta< \lfloor \delta_C/2 \rfloor$ conclude that $A_u$ contains a square $\mathscr S$ with horizontal and vertical edges of length $\lfloor \delta_C/2 \rfloor$: choose the center of $\mathscr S$ as $(c_u,\delta_C)$. In fact, the right corners of $\mathscr S$ may only be adjacent to $A_u$. Nevertheless, $\mathscr S \cap S_{\lfloor \delta_C/2 \rfloor}\not=\emptyset$, and thus $\mathscr S \cap S_{g}\not=\emptyset$ for $g\le \lfloor \delta_C/2 \rfloor$. \hfill $\Box$\\
	
	\noindent\textbf{Proof of Theorem \ref{theo:main}:} 
	For $(t_s,h_s)\in S$ set $\hat t_{e_s}^{(n)}  := \hat t_{e}^{(n)}$, making the relation to $(t_s,h_s)\in S$ explicit. Let $C\not=\emptyset$. Then a.s.~for $n$ large enough it is
	\begin{align}\label{maxmin}
	\max_{(t_s,h_s)\in S\backslash R} \Big(\min\{|\hat t_{e_s}^{(n)} - c_u|: c_u\in C\} \Big)\le \delta-1, 
	\end{align}
	i.e., all potential paths that start in the 'upper part' $S\backslash R$ of $\Delta_\delta$ end close to a $c_u\in C$. (\ref{maxmin}) holds true, as for any $(t_s,h_s)\in S\backslash R$ the path eventually enters an $A_u$ and then Proposition \ref{lemm:pathDth} states a.s.~$|\hat t_{e_s}^{(n)}-c_u|\le  \delta-1$ for $n$ large. The maximum follows from $|S|$ being finite. Further, a.s.~for $n$ large it is
	\begin{align}\label{minmax}
	\min_{(t_s,h_s)\in \bigcup_{u=1,\ldots,|C|} A_u}  
	[(nh)^{-1/2}\cdot|D_{t,h}^{(n)}|] > \max_{(t_s,h_s)\in R}  [(nh)^{-1/2}\cdot|D_{t,h}^{(n)}|],
	\end{align}
	i.e., as long as the algorithm does not break, starting points are first chosen from $S\backslash R\supset \bigcup_u A_u$ up until all of them are cut out, and after that they are chosen from $R$. (\ref{minmax}) follows from Corollary \ref{conv:Dth} and Lemma \ref{lemm_shark}, noting that $d_{t,h}^{(1)} = 0$ if $(t,h)\in R$, and  $|d_{t,h}^{(1)}| > 0$ if $(t,h)\in A_u$.
	
	Combining (\ref{maxmin}) and (\ref{minmax}), it follows that a.s.~for $n$ large at first all $c\in C$ are estimated up to a distance $ \delta -1$ from starting values within $S\backslash R$, given the algorithm does not break. Possibly step 3a is applied in between. We need to show that the breaking criterion 3b applies appropriately:
	First, for $m\le|C|$ we show that it does not apply. A path starting in $S\backslash R$ must pass an $A_u$ such that a.s.~for $n$ large
	\[
	\min_{(t_s,h_s)\in S\backslash R}
	\big[
	\max_{k=0,1\ldots, h_{s}-\delta} |D_{\hat t_{s}^{(n)}(k),h_{s}-k}^{(n)}| \big] 
	\ge \min_{(t,h)\in \bigcup A_u} |D_{t,h}^{(n)}| > n^\beta,
	\]
	i.e., the maximum w.r.t.~any path starting in $S\backslash R$ exceeds the minimum w.r.t.~to all $A_u$, and the second inequality holds as $D_{t,h}^{(n)}=n^{1/2} d_{t,h}^{(1)}+o_{a.s.}(n^{1/2})$ uniformly on $\Delta_\delta$ (Corollary \ref{conv:Dth}) and $|d_{t,h}^{(1)}|>0$ within all $A_u$ (Lemma \ref{lemm_shark}), noting that $\beta<1/2$. Thus, a.s. for $n$ large there is no break. 
	
	Second, for $m=|C|+1$ the algorithm now breaks. All $c\in C$ are estimated. Thus, possibly after applying criterion 3a, all remaining starting points lie in $R$. This also covers $C=\emptyset$. Note that a path that starts in $R$ remains in $R$. It holds a.s.~for $n$ large
	\[
	\max_{(t_s,h_s)\in R}
	\big[
	\max_{k=0,1\ldots, h_{s}-\delta} |D_{\hat t_{s}^{(n)}(k),h_{s}-k}^{(n)}| \big] 
	\le \max_{(t,h)\in R} |D_{t,h}^{(n)}| < n^{1/2-v},
	\]
	i.e., the maximum w.r.t.~all paths starting in $R$ deceeds the maximum w.r.t~the entire $R$. For the second inequality we note that within $R$ it is $d_{t,h}^{(1)}=0$ and thus $D_{t,h}^{(n)}=o_{a.s.}(n^{1/2-v})$ uniformly on $R$ (Corollary \ref{conv:Dth}). As $\beta\ge 1/2-v$, a.s.~for $n$ large the algorithm breaks. 
	
	Finally note that 3c.~is asymptotically redundant: a.s.~for $n$ large, for $m\le |C|$ we stated correct estimation up to the error, so neighboring estimates will have at least distance $\delta_C - 2( \delta -1)$, and for $m=|C|+1$ the algorithm already broke in 3b. \hfill $\Box$
	

\vspace{2cm}
	 \begin{small}
	 \bibliography{literature_arxiv}   

\begin{thebibliography}{}

\bibitem[Antoch and Hu\v{s}kov\'{a}, 1999]{Antoch1999}
Antoch, J. and Hu\v{s}kov\'{a}, M. (1999).
\newblock Estimators of changes.
\newblock In {\em Asymptotics, nonparametrics, and time series}, volume 158 of
  {\em Statist. Textbooks Monogr.}, pages 533--577. Dekker, New York.

\bibitem[Aston and Kirch, 2012]{Aston2012}
Aston, J. A.~D. and Kirch, C. (2012).
\newblock Evaluating stationarity via change-point alternatives with
  applications to fmri data.
\newblock {\em Ann. Appl. Stat.}, 6(4):1906--1948.

\bibitem[Aue and Horv{\'a}th, 2013]{Aue2013}
Aue, A. and Horv{\'a}th, L. (2013).
\newblock Structural breaks in time series.
\newblock {\em J. Time Ser. Anal.}, 34(1):1--16.

\bibitem[Baranowski et~al., 2019a]{Baranowski2019}
Baranowski, R., Chen, Y., and Fryzlewicz, P. (2019a).
\newblock Narrowest-over-threshold detection of multiple change-points and
  change-point-like features.
\newblock {\em J. R. Stat. Soc., B: Stat. Methodol.}, 81:649--672.

\bibitem[Baranowski et~al., 2019b]{CRANnot}
Baranowski, R., Chen, Y., and Fryzlewicz, P. (2019b).
\newblock {\em not: Narrowest-Over-Threshold Change-Point Detection}.
\newblock R package version 1.2.

\bibitem[Baranowski and Fryzlewicz, 2019]{CRANwbs}
Baranowski, R. and Fryzlewicz, P. (2019).
\newblock {\em wbs: Wild Binary Segmentation for Multiple Change-Point
  Detection}.
\newblock R package version 1.4.

\bibitem[Berkes et~al., 2006]{Berkes2006}
Berkes, I., Horv{\'a}th, L., Kokoszka, P., and Shao, Q.-M. (2006).
\newblock On discriminating between long-range dependence and changes in mean.
\newblock {\em Ann. Statist.}, 34(3):1140--1165.

\bibitem[Brodsky, 2017]{Brodsky2017}
Brodsky, B. (2017).
\newblock {\em Change-point analysis in nonstationary stochastic models}.
\newblock CRC Press, Boca Raton, FL.

\bibitem[Chen and Gupta, 2000]{Chen2000}
Chen, J. and Gupta, A.~K. (2000).
\newblock {\em Parametric statistical change point analysis}.
\newblock Birkh\"{a}user Boston, Inc., Boston, MA.

\bibitem[Cho and Kirch, 2022]{Cho2020}
Cho, H. and Kirch, C. (2022).
\newblock Two-stage data segmentation permitting multiscale change points,
  heavy tails and dependence.
\newblock {\em Ann Inst Stat Math}, 74(4):653--684.

\bibitem[Chu et~al., 1995]{Chu1995}
Chu, C.-S.~J., Hornik, K., and Kuan, C.-M. (1995).
\newblock M{OSUM} tests for parameter constancy.
\newblock {\em Biometrika}, 82(3):603--617.

\bibitem[Cs\"{o}rg\H{o} and Horv\'{a}th, 1997]{Csorgo1997}
Cs\"{o}rg\H{o}, M. and Horv\'{a}th, L. (1997).
\newblock {\em Limit theorems in change-point analysis}.
\newblock Wiley Series in Probability and Statistics. John Wiley \& Sons, Ltd.,
  Chichester.
\newblock With a foreword by David Kendall.

\bibitem[Dehling et~al., 2017]{Dehling2017}
Dehling, H., Rooch, A., and Taqqu, M.~S. (2017).
\newblock Power of change-point tests for long-range dependent data.
\newblock {\em Electron. J. Stat.}, 11(1):2168--2198.

\bibitem[Dette et~al., 2020]{Dette2020}
Dette, H., Eckle, T., and Vetter, M. (2020).
\newblock Multiscale change point detection for dependent data.
\newblock {\em Scand. J. Stat.}, 47(4):1243--1274.

\bibitem[D\"{o}ring, 2010]{Doering2010}
D\"{o}ring, M. (2010).
\newblock Multiple change-point estimation with {$U$}-statistics.
\newblock {\em J. Statist. Plann. Inference}, 140(7):2003--2017.

\bibitem[Eichinger and Kirch, 2018]{Eichinger2018}
Eichinger, B. and Kirch, C. (2018).
\newblock A mosum procedure for the estimation of multiple random change
  points.
\newblock {\em Bernoulli}, 24(1):526--564.

\bibitem[Fang et~al., 2020]{Fang2020}
Fang, X., Li, J., and Siegmund, D. (2020).
\newblock Segmentation and estimation of change-point models: false positive
  control and confidence regions.
\newblock {\em Ann. Statist.}, 48(3):1615--1647.

\bibitem[Fryzlewicz, 2014]{fryz2014}
Fryzlewicz, P. (2014).
\newblock Wild binary segmentation for multiple change-point-detection.
\newblock {\em Ann. Statist.}, 42(6):2243--2281.

\bibitem[Fryzlewicz, 2018a]{Fryzlewicz2018s}
Fryzlewicz, P. (2018a).
\newblock Supplement to "tail-greedy bottom-up data decompositions and fast
  multiple change-point detection".

\bibitem[Fryzlewicz, 2018b]{Fryzlewicz2018p}
Fryzlewicz, P. (2018b).
\newblock Tail-greedy bottom-up data decompositions and fast multiple
  change-point detection.
\newblock {\em Ann. Statist.}, 46(6B):3390--3421.

\bibitem[Gombay and Horv\'{a}th, 1994]{Gombay1994}
Gombay, E. and Horv\'{a}th, L. (1994).
\newblock An application of the maximum likelihood test to the change-point
  problem.
\newblock {\em Stochastic Process. Appl.}, 50(1):161--171.

\bibitem[Gombay and Horv\'{a}th, 2002]{Gombay2002}
Gombay, E. and Horv\'{a}th, L. (2002).
\newblock Rates of convergence for {$U$}-statistic processes and their
  bootstrapped versions.
\newblock volume 102, pages 247--272.
\newblock Silver jubilee issue.

\bibitem[Harchaoui and L\'{e}vy-Leduc, 2010]{Harchaoui2010}
Harchaoui, Z. and L\'{e}vy-Leduc, C. (2010).
\newblock Multiple change-point estimation with a total variation penalty.
\newblock {\em J. Amer. Statist. Assoc.}, 105(492):1480--1493.

\bibitem[Hinkley, 1971]{Hinkley1971}
Hinkley, D.~V. (1971).
\newblock Inference about the change-point from cumulative sum tests.
\newblock {\em Biometrika}, 58(3):509--523.

\bibitem[Holmes et~al., 2013]{Holmes2013}
Holmes, M., Kojadinovic, I., and Quessy, J.-F. (2013).
\newblock Nonparametric tests for change-point detection \`a la {G}ombay and
  {H}orv\'{a}th.
\newblock {\em J. Multivar. Anal.}, 115:16--32.

\bibitem[Horv\'{a}th and Hu\v{s}kov\'{a}, 2005]{Horvath2005}
Horv\'{a}th, L. and Hu\v{s}kov\'{a}, M. (2005).
\newblock Testing for changes using permutations of {U}-statistics.
\newblock {\em J. Statist. Plann. Inference}, 128(2):351--371.

\bibitem[Horv\'{a}th and Shao, 2007]{Horvath2007}
Horv\'{a}th, L. and Shao, Q.-M. (2007).
\newblock Limit theorems for permutations of empirical processes with
  applications to change point analysis.
\newblock {\em Stochastic Process. Appl.}, 117(12):1870--1888.

\bibitem[Hu\v{s}kov\'{a} and Slab\'{y}, 2001]{Huskova2001}
Hu\v{s}kov\'{a}, M. and Slab\'{y}, A. (2001).
\newblock Permutation tests for multiple changes.
\newblock {\em Kybernetika}, 37(5):605--622.

\bibitem[Kass-Hout et~al., 2012]{Kass2012}
Kass-Hout, T., Xu, Z., Mcmurray, P., Park, S., Buckeridge, D., Brownstein, J.,
  Finelli, L., and Groseclose, S. (2012).
\newblock Application of change point analysis to daily influenza-like illness
  emergency department visits.
\newblock {\em JAMIA}, 19:1075--81.

\bibitem[Killick et~al., 2010]{Killick2010}
Killick, R., Eckley, I.~A., Ewans, K., and Jonathan, P. (2010).
\newblock Detection of changes in variance of oceanographic time-series using
  changepoint analysis.
\newblock {\em Ocean Eng.}, 37(13):1120--1126.

\bibitem[Killick et~al., 2016]{CRANchangepoint}
Killick, R., Haynes, K., and Eckley, I.~A. (2016).
\newblock {\em {changepoint}: An {R} package for changepoint analysis}.
\newblock R package version 2.2.2.

\bibitem[Lavielle and Moulines, 2000]{Lavielle2000}
Lavielle, M. and Moulines, E. (2000).
\newblock Least-squares estimation of an unknown number of shifts in a time
  series.
\newblock {\em J. Time Ser. Anal.}, 21(1):33--59.

\bibitem[Levajkovi\'c and Messer, 2021]{CRANMSCP}
Levajkovi\'c, T. and Messer, M. (2021).
\newblock {\em mscp: Multiscale Change Point Detection via Gradual Bandwidth
  Adjustment in Moving Sum Processes}.
\newblock R package version 1.0.

\bibitem[Li and Sieling, 2017]{CRANFDRSeg}
Li, H. and Sieling, H. (2017).
\newblock {\em FDRSeg: FDR-Control in Multiscale Change-Point Segmentation}.
\newblock R package version 1.0-3.

\bibitem[Matteson and James, 2014]{matteson2014}
Matteson, D.~S. and James, N.~A. (2014).
\newblock A nonparametric approach for multiple change point analysis of
  multivariate data.
\newblock {\em J. Am. Stat. Assoc.}, 109(505):334--345.

\bibitem[Meier et~al., 2021]{CRANmosum}
Meier, A., Kirch, C., and Cho, H. (2021).
\newblock {mosum}: A package for moving sums in change-point analysis.
\newblock {\em J. Stat. Softw.}, 97(8):1--42.

\bibitem[Messer, 2022]{Messer2021}
Messer, M. (2022).
\newblock Bivariate change point detection: joint detection of changes in
  expectation and variance.
\newblock {\em Scand. J. Stat.}, 49:886--916.

\bibitem[Messer et~al., 2014]{Messer2014}
Messer, M., Kirchner, M., Schiemann, J., Roeper, J., Neininger, R., and
  Schneider, G. (2014).
\newblock A multiple filter test for the detection of rate changes in renewal
  processes with varying variance.
\newblock {\em Ann. Appl. Stat.}, 8(4):2027--2067.

\bibitem[Muggeo, 2020]{CRANcumSeg}
Muggeo, V.~M. (2020).
\newblock {\em cumSeg: Change Point Detection in Genomic Sequences}.
\newblock R package version 1.3.

\bibitem[Nurk et~al., 2022]{Nurk2022}
Nurk, S., Koren, S., Rhie, A., Rautiainen, M., Bzikadze, A., Mikheenko, A.,
  Vollger, M., Altemose, N., Uralsky, L., Gershman, A., Aganezov, S., Hoyt, S.,
  Diekhans, M., Logsdon, G., Alonge, M., Antonarakis, S., Borchers, M.,
  Bouffard, G., Brooks, S., Caldas, G., Chen, N., Cheng, H., Chin, C., Chow,
  W., de~Lima, L., Dishuck, P., Durbin, R., Dvorkina, T., Fiddes, I., Formenti,
  G., Fulton, R., Fungtammasan, A., Garrison, E., Grady, P., Graves-Lindsay,
  T., Hall, I., N.F., H., Hartley, G., Haukness, M., Howe, K., Hunkapiller, M.,
  Jain, C., Jain, M., Jarvis, E., Kerpedjiev, P., Kirsche, M., Kolmogorov, M.,
  Korlach, J., Kremitzki, M., Li, H., Maduro, V., Marschall, T., McCartney, A.,
  McDaniel, J., Miller, D., Mullikin, J., Myers, E., Olson, N., Paten, B.,
  Peluso, P., Pevzner, P., Porubsky, D., Potapova, T., Rogaev, E., Rosenfeld,
  J., Salzberg, S., Schneider, V., Sedlazeck, F., Shafin, K., Shew, C.,
  Shumate, A., Sims, Y., Smit, A., Soto, D., Sovi\'~c, I., Storer, J., Streets,
  A., Sullivan, B., Thibaud-Nissen, F., Torrance, J., Wagner, J., Walenz, B.,
  Wenger, A., Wood, J., Xiao, C., Yan, S., Young, A., Zarate, S., Surti, U.,
  McCoy, R., Dennis, M., Alexandrov, I., Gerton, J., O'Neill, R., Timp, W.,
  Zook, J., Schatz, M., Eichler, E., Miga, K., and Phillippy, A. (2022).
\newblock The complete sequence of a human genome.
\newblock {\em Science}, 376:44--53.

\bibitem[Page, 1954]{Page1954}
Page, E.~S. (1954).
\newblock Continuous inspection schemes.
\newblock {\em Biometrika}, 41(1-2):100--115.

\bibitem[Pein et~al., 2020]{CRANstepR}
Pein, F., Hotz, T., Sieling, H., and Aspelmeier, T. (2020).
\newblock {\em {stepR}: Multiscale change-point inference}.
\newblock R package version 2.1-1.

\bibitem[Pein et~al., 2017]{Pein2016}
Pein, F., Sieling, H., and Munk, A. (2017).
\newblock Heterogeneous change point inference.
\newblock {\em J. R. Stat. Soc. Ser. B Methodol.}, 79(4):1207--1227.

\bibitem[Reeves et~al., 2007]{Reeves2007}
Reeves, J., Chen, J., Wang, X.~L., Lund, R., and Lu, Q.~Q. (01 Jun. 2007).
\newblock A review and comparison of changepoint detection techniques for
  climate data.
\newblock {\em JAMC}, 46(6):900--915.

\bibitem[{Rybach} et~al., 2009]{Rybach2009}
{Rybach}, D., {Gollan}, C., {Schluter}, R., and {Ney}, H. (2009).
\newblock Audio segmentation for speech recognition using segment features.
\newblock In {\em 2009 IEEE International Conference on Acoustics, Speech and
  Signal Processing}, pages 4197--4200.

\bibitem[Spokoiny, 2009]{Spokoiny2009}
Spokoiny, V. (2009).
\newblock Multiscale local change point detection with applications to
  value-at-risk.
\newblock {\em Ann. Statist.}, 37(3):1405--1436.

\bibitem[Steinebach and Eastwood, 1995]{Steinebach1995}
Steinebach, J. and Eastwood, V.~R. (1995).
\newblock On extreme value asymptotics for increments of renewal processes.
\newblock volume~45, pages 301--312.
\newblock Extreme value theory and applications (Villeneuve d'Ascq, 1992).

\end{thebibliography}
	 \end{small}
	 
	 \vfill
	 \section*{Contact information}
	 Tijana Levajkovi\'c\\
	 email.: \href{mailto:tijana.levajkovic@tuwien.ac.at}{tijana.levajkovic@tuwien.ac.at}\\
	 Michael Messer\\
	 email.: \href{mailto:michael.messer@tuwien.ac.at}{michael.messer@tuwien.ac.at}
	 
	 \vspace{0.5em}\noindent
	 Vienna University of Technology\\ 
	 Institute of Statistics and Mathematical Methods in Economics\\ 
	 Wiedner Hauptstraße 8-10/105\\ 
	 1040 Vienna, Austria 
	 
	\end{document}